\documentclass[%
 pra,
 twocolumn,
 amsmath,amssymb,groupaddress,
 aps,
]{revtex4-2}
\usepackage[colorlinks,allcolors=blue]{hyperref}

\usepackage{physics,xcolor,palatino}
\usepackage{graphicx}
\usepackage{dcolumn}
\usepackage{bm}
\usepackage{ulem}

\graphicspath{ {FigsForDraft/} }

\begin{document}


\title{Engineering nonlinear boson-boson interactions using mediating spin systems}

\author{Hannah McAleese}
\affiliation{Centre for Quantum Materials and Technologies, School of Mathematics and Physics, Queen’s University Belfast, BT7 1NN Belfast, UK}

\author{Mauro Paternostro}
\affiliation{Universit\`a degli Studi di Palermo, Dipartimento di Fisica e Chimica - Emilio Segr\`e, via Archirafi 36, I-90123 Palermo, Italy}
\affiliation{Centre for Quantum Materials and Technologies, School of Mathematics and Physics, Queen’s University Belfast, BT7 1NN Belfast, UK}

\author{Ricardo Puebla}
\affiliation{Departamento de F\'isica, Universidad Carlos III de Madrid, Avda. de la Universidad 30, 28911 Legan\'es, Spain}

\date{\today}

\begin{abstract}
We present a protocol to create entangled coherent states by engineering cross-Kerr interactions between bosonic systems endowed with (externally driven) internal spin-like degrees of freedom. With slight modifications, the protocol is also able to produce N00N states through nonlinear beam splitter interactions. {Each bosonic system interacts locally with its spin and by suitably tuning the model parameters, various classes of effective bosonic interaction Hamiltonians, mediated by the coupled spins, can be engineered.} Our approach is benchmarked by numerical simulations aimed at studying the entanglement within a bosonic register and comparing it with the expected one resulting from the target Hamiltonians. 
\end{abstract}

\maketitle

Entangled bosonic states are valuable resources in quantum computation, communication and  sensing. For instance, entangled coherent states (ECSs)~\cite{Sanders2012} can be used for quantum communication applications such as entanglement distribution~\cite{Sangouard2010,Brask2010,Lim2016} and quantum teleportation~\cite{vanEnk2001,Wang2001,Jeong2001,Joo2016}. They are equally key for continuous-variable quantum computation where coherent states are logical qubits. ECSs also form part of continuous-variable gate operations, such as CNOT gates~\cite{Jeong2002}. As a second remarkable example, N00N states are notable for their applications in quantum metrology and quantum imaging, due to their ability for super-resolution and super-sensitivity~\cite{Boto2000,DAngelo2001,Kok2004,Nagata2007,Dowling2008,Israel2014,Slussarenko2017}. 

The price to pay for such a wide range of topical applications is the need for non-linear processes to engineer such states. As an example, an ECS having the form 
\begin{equation}
    \ket{\mathrm{ECS}} \propto \ket{\alpha}_1 \ket{\alpha}_2 + \ket{-\alpha}_1 \ket{-\alpha}_2
\end{equation}
with $\ket{\alpha}_j$ a coherent state of boson $j=1,2$ with amplitude $\alpha\in\mathbb{C}$, would require a cross-Kerr interaction between bosons 1 and 2, while the N00N state
\begin{equation}
    \ket{\mathrm{N00N}} \propto \ket{N}_1 \ket{0}_2 + \ket{0}_1 \ket{N}_2
\end{equation}
would necessitate non-linear beam-splitter interactions, were we to produce them in a direct way. Natural non-linear mechanisms of this sort are typically very inefficient in light of weak rates of non-linearities, which makes it necessary to seek usually convoluted and resource-expensive Hamiltonian-engineering schemes. It is thus appealing to investigate ways to overcome these difficulties and achieve sufficiently large rates of non-linearity and the desired operatorial form of the interaction Hamiltonian, with only a modest resource overhead.

In this paper, we propose a method for the synthesis of both ECSs and N00N states inspired by the results reported in Ref.~\cite{Casanova2018}. {There, it was shown} that linear spin-boson couplings and spin drivings are sufficient to approximately manufacture nonlinear spin-boson interactions in the form of an $n$th order quantum Rabi model. Taking issue from this, here we address the question: "{\it Would such an approach be successful in achieving non-linear bosonic interactions when considering more than a single boson?"}. We pursue this question along two different directions. First, {we aim to design a deterministic protocol for the engineering of ECSs } having large coherent amplitudes. For $\alpha\gtrsim1$, in fact, we have $\langle-\alpha\vert\alpha\rangle\simeq0$, which guarantees the nearly perfect distinguishability of the components of $\ket{\mathrm{ECS}}$ and thus maximizes its degree of entanglement. ECSs with coherent state amplitudes of $|\alpha|=0.65$ have been achieved through photon subtraction~\cite{Ourjoumtsev2009}, though this method is probabilistic and requires the creation of Schr{\"{o}}dinger cat states of the form $\ket{\alpha}+\ket{-\alpha}$, which is a difficult task {\it per se}. Approximate ECSs with $|\alpha|\approx 0.39$ have been produced using a deterministic method of mixing coherent and squeezed light~\cite{Israel2019}. Using an alternative source of squeezed light should allow for amplitudes up to $|\alpha|=1$ using the same technique~\cite{Israel2019}, although this has not yet been achieved. Ref.~\cite{Wang2016} has reached $\alpha=1.92$ experimentally through a deterministic process with microwave cavities interacting via a superconducting qubit. A similar proposal for magnons and phonons interacting through a transmon qubit~\cite{Kounalakis2023} takes amplitudes of just over $\alpha=2$. However, these restricted setups make it impossible for the engineered ECS to be used in long-haul communication processes and for distributed quantum computation. ECSs of an alternative form $\ket{\alpha}\ket{\beta} + \ket{-\alpha}\ket{-\beta}$ were created in the motional modes of a trapped ion~\cite{Jeon2024} where the amplitudes reached $|\beta|=2$ and $|\alpha|=0.7$. The method we present in this work is not limited to any one experimental setting and can successfully entangle coherent states with amplitudes of $\alpha>2$.

Many methods have been proposed for producing N00N states (see Refs.~\cite{Gerry2001,Kok2002,Pryde2003,Walther2004,Mitchell2004,Resch2007,Cable2007,Hofmann2007,Jones2009,Afek2010,Merkel2010,Wang2011,Zhang2018} for some examples). Thus far, N00N states with up to $N=5$ have been achieved in photonic systems~\cite{Afek2010}, $N=9$ using the vibrational modes of a trapped ion~\cite{Zhang2018}, and $N=10$ in a spin system~\cite{Jones2009}. While our proposal is limited in the size of N00N states it can produce, it has the advantage of added flexibility; the second direction that we pursue in this work is that of devising an effective protocol -- based on nonlinear spin-boson couplings -- that naturally allows for the creation of (un)even N00N states, also referred to as N00M or M00N states~\cite{Wang2011,ZunigaSegundo2013}, which consist of the superposition of $\ket{N}_1\ket{0}_2$ and $\ket{0}_1\ket{M}_2$ with $M\neq N$. Strategies to create these states have been devised for superconducting microwave resonators~\cite{Wang2011} and trapped ions~\cite{ZunigaSegundo2013}, but we put forward a general method which is not restricted to one particular physical system.

The remainder of this article is structured as follows. In Sec.~\ref{sec:spinBosonSystem}, we show how to manipulate the dynamics of one spin-boson system with linear couplings in order to create effective non-linear dynamics. This result proves useful in the rest of our analysis. We focus on replicating a cross-Kerr Hamiltonian of two {bosonic modes} in Sec.~\ref{sec:crossKerr}; firstly, in Sec.~\ref{sec:deriveCrossKerr} we find a protocol to generate this interaction. Secondly, in Sec.~\ref{sec:resultsCrossKerr} we present the results of numerical simulations comparing the effective dynamics entailed by the engineered non-linear model to the predictions stemming from the exact target Hamiltonian. We then shift our emphasis to fabricating N00M states via a non-linear beam splitter interaction in Sec.~\ref{sec:NOON}, demonstrating the method needed in Sec.~\ref{sec:deriveNOON} and analyzing the corresponding performance  in Sec.~\ref{sec:resultsNOON}. We lastly draw our conclusions in Sec.~\ref{sec:conc}.

\section{Spin-boson system} \label{sec:spinBosonSystem}

\begin{figure}[t]
\centering
\includegraphics[width=0.9\columnwidth]{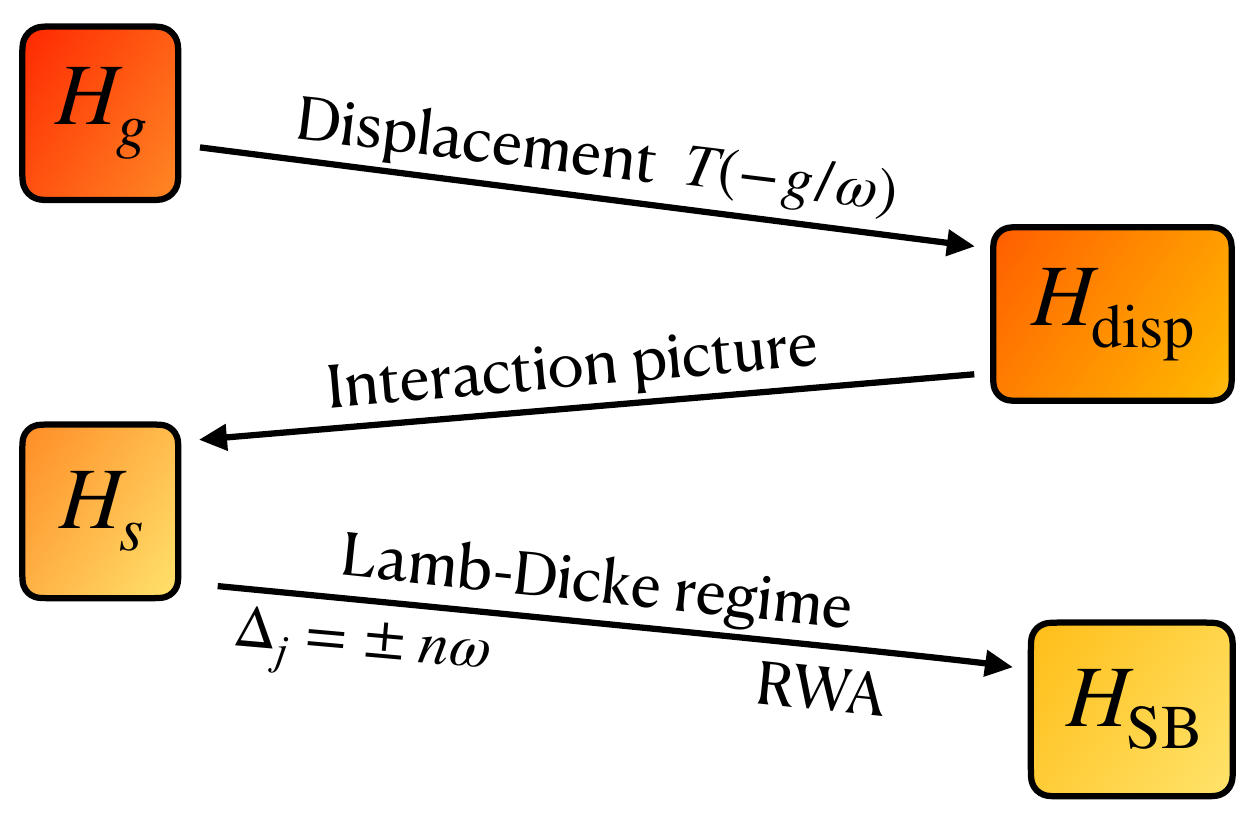}
\caption{Schematic diagram showing the steps needed to derive the Hamiltonian in Eq.~\eqref{eq:HamAfterApprox}.}
\label{fig:deriveHamiltonian}
\end{figure}

We begin by describing a driven single spin-boson system where the spin is subject to a number $n_d$ of rotations. For the $j$th rotation, we denote the driving frequency $\Delta_j$ and phase $\phi_j$. The bosonic mode has frequency $\omega$ and annihilation and creation operators $a$ and $a^\dagger$ respectively. The system acts according to the Hamiltonian {(unless otherwise specified,  throughout the manuscript we will use units such that $\hbar=1$)}
\begin{equation} \label{eq:linearHam}
    H_\mathrm{g} =  H_0 + \sum_{j=0}^{n_d} \frac{\epsilon_j}{2} [\cos (\Delta_j t + \phi_j) \sigma^z + \sin (\Delta_j t + \phi_j)\sigma^y],
\end{equation}
where $H_0=\omega a^\dagger a + g \sigma^x (a + a^\dagger)$ with $\sigma^{k}$ the $k=x,y,z$ Pauli matrix acting on the spin and $g$ the coupling strength of the linear spin-boson interaction. Finally, { $\epsilon_j$ refers to the strength of the driving mechanism acting on the spin itself.} 

Following the work in Refs.~\cite{Casanova2018,Puebla2019}, we show how Eq.~\eqref{eq:linearHam} is approximately equivalent to a Hamiltonian with nonlinear spin-boson interactions. An outline of the required stages of this process is displayed in Fig.~\ref{fig:deriveHamiltonian}. We first act on $H_\mathrm{g}$ with a spin-dependent displacement operator $T(\alpha)$, in order to introduce higher order terms of the bosonic operators into the Hamiltonian. This can be written in the spin basis $\{\ket{g},\ket{e}\}$ (with $\sigma^z = \ket{e}\bra{e} - \ket{g}\bra{g}$) as
\begin{equation}
    T(\alpha) = \frac{1}{\sqrt{2}} \left(
    \begin{array}{cc}
        D^\dagger (\alpha) & D(\alpha) \\
        -D^\dagger (\alpha) & D(\alpha) 
    \end{array}
    \right),
\end{equation}
where $D(\alpha) = e^{\alpha a^\dagger - \alpha^* a}$ is the displacement operator of amplitude $\alpha$ acting on the bosonic mode. Transforming $H_\mathrm{g}$ using a displacement amplitude of $-g/\omega$, we obtain 
\begin{equation} \label{eq:dispHam}
\begin{split}
    H_\mathrm{disp} & = T^\dagger (-g/\omega) H_\mathrm{g} T(-g/\omega) \\ & = \omega a^\dagger a + \sum_{j=0}^{n_d} \frac{\epsilon_j}{2} \Big\{ \sigma^+ e^{\eta(a-a^\dagger)} e^{-i(\Delta_j t +\phi_j)} + \mathrm{h.c.} \Big\},
\end{split}
\end{equation}
where $\eta=2g/\omega$ and $\sigma^+=\ket{e}\bra{g}$. We now have nonlinear spin-boson interaction terms coming from the new $\sigma^+ e^{\eta(a-a^\dagger)}$ term, which we will later leverage to generate nonlinear coupling between two bosons. We next move to an interaction picture with respect to  $\omega a^\dagger a$, resulting in the Hamiltonian
\begin{equation} \label{eq:sHam}
    H_s = \sum_{j=0}^{n_d} \frac{\epsilon_j}{2} \left\{ \sigma^+ e^{\eta(a(t)-a^\dagger (t))} e^{-i(\Delta_j t +\phi_j)} + \mathrm{h.c.} \right\},
\end{equation}
where $a(t) = a e^{-i\omega t}$.

We now need to make some approximations, which will require imposing constraints on the parameters in our system. Firstly, we tune the driving frequencies $\Delta_j$ to $\pm n\omega $ ($n \in \mathbb{Z}^+$). Secondly, we require $|\Delta_j| \gg |\epsilon_j|$ so as to claim for the rotating wave approximation (RWA)  that will allow us to discard the rapidly oscillating terms $e^{-i \Delta_j t}$ in $H_\mathrm{s}$. Thirdly, we invoke the Lamb-Dicke regime by requiring $|\eta|\sqrt{\langle (a + a^\dagger)^2 \rangle} \ll 1$, which enables us to neglect high-order terms in the expansion of $e^{\eta(a(t)-a^\dagger (t))}$. This set of  approximations result in the effective non-linear spin-boson Hamiltonian
\begin{equation} \label{eq:HamAfterApprox}
    H_{\mathrm{SB}} = \sum_j \frac{\epsilon_j}{2} \left[ \sigma^+ e^{-i \phi_j} f(\Delta_j) + \mathrm{h.c.} \right],
\end{equation}
where
\begin{equation}
    f(\Delta_j) = 
    \begin{cases}
        \dfrac{\eta^n}{n!} (-a^\dagger)^n, & \text{if $\Delta_j = +n\omega$,} \\
        \noalign{\vskip9pt}
        \dfrac{\eta^n}{n!} (a^n), & \text{if $\Delta_j = -n\omega$,} \\
        \noalign{\vskip9pt}
        1-\dfrac{\eta^2}{2} -\eta^2 a^\dagger a, & \text{if $\Delta_j = 0$.}
    \end{cases}
\end{equation}

\subsection{Two spin-boson systems}

Having analyzed how a single spin-boson system is transformed under the described map, we now extend it to a system made of two subsystems consisting of a spin and a bosonic mode. That is, the full Hamiltonian can be written as $H_f=H_{\mathrm{g}1}+H_{\mathrm{g}2}+H_{int}$ where $H_{\mathrm{g}1}$ and $H_{\mathrm{g}2}$ is given by Eq.~\eqref{eq:linearHam} for each single spin-boson subsystem, while $H_{int}$ refers to a interaction between the spins 1 and $2$. We further assume that the interaction only couples spins and is of the form $H_{int}=\lambda \sigma_1^x\sigma_2^x$. This physical interaction between spins is standard in many quantum setups, such as in superconducting qubits or trapped ions, where the inter-spin coupling strength $\lambda$ depends on the specific implementation.  Noting that $T^\dagger(\alpha)\sigma_xT(\alpha)=-\sigma_z$, the total Hamiltonian $H_f$ transforms into
\begin{equation} \label{eq:totalHamiltonian}
    H = H_{\mathrm{SB}1} + H_{\mathrm{SB}2} + \lambda \sigma_1^z \sigma_2^z,
\end{equation}
where $H_{\mathrm{SB}j}$ ($j=1,2$) reads as in Eq.~\eqref{eq:HamAfterApprox} for the $j$th non-linear spin-boson subsystem, and $\lambda$ is the inter-spin coupling strength. 

\section{Engineering a Cross-Kerr Interaction} \label{sec:crossKerr}

\subsection{Developing the Protocol} \label{sec:deriveCrossKerr}

Our first goal is to simulate a cross-Kerr interaction, which is the crucial ingredient in the task of engineering an ECS. We begin by taking one driving term with $\Delta=0$ and $\phi=0$ in the system in Eq.~(\ref{eq:totalHamiltonian}). {In this case, the Hamiltonian can be approximated by}
\begin{equation} \label{eq:hamBeforeMagnusCK}
    H  = \sum^2_{j=1}\frac{\epsilon_j}{2} \sigma_j^x \left( 1 - \frac{\eta_j^2}{2} - \eta_j^2 a^\dagger_j a_j \right) + \lambda \sigma_1^z \sigma_2^z,
\end{equation}
{where rotating terms at frequencies $\pm n\omega$ have been neglected and we have labelled as} $a_j$ the annihilation operator of the bosonic mode of subsystem $j=1,2$. We then move to an interaction picture with respect to $H_\mathrm{spins}=\lambda \sigma_1^z \sigma_2^z$ to find the Hamiltonian $H_I$, which we can express in the basis $\mathcal{S}_{CK} = \{ \ket{\phi_1}, \ket{\phi_2}, \ket{\phi_3}, \ket{\phi_4} \}$ where
\begin{equation} \label{eq:basisCK}
\begin{split}
    & \ket{\phi_1} = \ket{e,p}_1 \otimes \ket{e,q}_2, \quad \ket{\phi_2} = \ket{e,p}_1 \otimes \ket{g,q}_2, \\ & \ket{\phi_3} = \ket{g,p}_1 \otimes \ket{e,q}_2, \quad \ket{\phi_4} = \ket{g,p}_1 \otimes \ket{g,q}_2
\end{split}
\end{equation}
with $\ket{p}_1$ and $\ket{q}_2$ number states of the first and second bosonic mode respectively.
\begin{widetext}
 The interaction Hamiltonian is then
\begin{equation} \label{eq:intHam}
\begin{split}
    H_I (t) &=  e^{i2\lambda t} \left[\tilde{g}_1 \left( \ket{\phi_1}\bra{\phi_3} + \ket{\phi_4}\bra{\phi_2} \right) + \tilde{g}_2 \left( \ket{\phi_1}\bra{\phi_2} + \ket{\phi_4}\bra{\phi_3} \right) \right]+ \mathrm{h.c.},
\end{split}
\end{equation}
where $\tilde{g}_1 = \epsilon_1 (1-\eta_1^2/2-\eta_1^2 p)/2$ and $\tilde{g}_2 = \epsilon_2 (1-\eta_2^2/2-\eta_2^2 q)/2$. 
\end{widetext}
Note that although the Hamiltonian~\eqref{eq:hamBeforeMagnusCK} is time independent, it can produce non-trivial dynamics in the bosonic degrees of freedom mediated by the spin interactions. In order to obtain the effective bosonic interactions produced by the time evolution, we use the Magnus expansion~\cite{Magnus1954} of the time-ordered time propagator $U_I (t) = \mathcal{T} \exp[-i\int^t_0 ds H_I (s)]$ where $\mathcal{T}$ is the Dyson time-ordering operator. Therefore, we consider $U_I (t) = \exp \left( \sum_{k=1}^\infty \Delta_k (t) \right)$ and focus on the first two terms of the series, which read
\begin{equation}
\begin{split}
    & \Delta_1 (t) = -i \int_0^t ds\, H_I (s), \\
    & \Delta_2 (t) = -\frac{1}{2} \int^t_0 ds_1 \int^{s_1}_0 ds_2 [H_I(s_1), H_I(s_2)].
\end{split}
\end{equation}
We find that, provided $\lambda$ is sufficiently large {with respect to the parameters $\epsilon_{1,2}$ or $\tilde{g}_{1,2}$}, it is possible to neglect all but one term from $\Delta_2 (t)$ which is proportional to $t$. The time evolution operator then becomes $U_I(t) = \exp(-it H_\mathrm{eff})$ where $H_\mathrm{eff}$ is a time-independent operator given by (in the $\mathcal{S}_{CK}$  basis)
\begin{equation} \label{eq:matrixHam}
H_\mathrm{eff} = \left(\begin{array}{cccc}
    \alpha & 0 & 0 & \beta \\
    0 & -\alpha & -\beta & 0 \\
    0 & -\beta & -\alpha & 0 \\
    \beta & 0 & 0 & \alpha
\end{array} \right),
\end{equation}
 and where we have introduced the parameters $\alpha = (\tilde{g}_1^2 + \tilde{g}_2^2)/(2\lambda)$ and $\beta = \tilde{g}_1 \tilde{g}_2 / \lambda$. 

The eigenstates of $H_\mathrm{eff}$ are $\{ (\ket{\phi_4} \pm \ket{\phi_1})/\sqrt{2}, (\ket{\phi_3} \pm \ket{\phi_2})/\sqrt{2} \}$ with corresponding eigenvalues $\{ \alpha \pm \beta, -\alpha \mp \beta \}$. Therefore, taking the initial state to be $\ket{\psi(0)} = (\ket{\phi_1} + \ket{\phi_4})/\sqrt{2}$, the evolved state at time $t$ is
\begin{equation}
    \ket{\psi(t)} =  e^{-i (\alpha + \beta + \lambda) t} \ket{\psi (0)}.
\end{equation}
This can be recast in terms of $p$ and $q$, neglecting immaterial global phases, as
\begin{equation} \label{eq:evolvedStatePhi1Phi4}
    \ket{\psi (t)} = e^{-it (\theta_p p + \theta_q q + \varphi_p p^2 + \varphi_q q^2 + \xi_{pq} pq)} \ket{\psi(0)},
\end{equation}
where 
\begin{align}
    & \theta_p = \frac{\epsilon_1 \eta_1^2 (\epsilon_1 (\eta_1^2 - 2) + \epsilon_2 (\eta_2^2 - 2))}{8 \lambda}, \\
    & \theta_q = \frac{\epsilon_2 \eta_2^2 (\epsilon_1 (\eta_1^2 - 2) + \epsilon_2 (\eta_2^2 - 2))}{8 \lambda}, \\
    & \varphi_p = \frac{\epsilon_1^2 \eta_1^4}{8 \lambda}, ~~\varphi_q = \frac{\epsilon_2^2 \eta_2^4}{8 \lambda},~~ \xi_{pq} = \frac{\epsilon_1 \epsilon_2 \eta_1^2 \eta_2^2}{4 \lambda} \label{eq:crossKerrCoeff}.
\end{align}
Upon tracing out the state of the spins, the two bosonic modes evolve according to the resulting model
\begin{equation} \label{eq:hamCK}
    H = \theta_p n_1 + \theta_q n_2 + \varphi_p n_1^2 + \varphi_q n_2^2 + \xi_{pq} n_1n_2
\end{equation}
with $n_j=a^\dag_ja_j$ the number operator of the $j$th mode. While the Hamiltonian includes the desired cross-Kerr term $n_1n_2$, it also encompasses self-Kerr ones $n_j^2$, which would compete with the cross-Kerr contribution and thus deplete the fidelity of the resultant states to the ECS we aim to generate~\cite{Joo2016}. We thus look for a protocol resulting in an effective Hamiltonian with no self-Kerr effect.

To achieve this, we notice that the choice of initial state $\ket{\psi(0)} = (-\ket{\phi_2}+\ket{\phi_3})/\sqrt{2}$ gives rise to the evolution
\begin{equation} \label{eq:evolvedStatePhi2Phi3}
    \ket{\psi(t)} \approx e^{-i t (\theta'_p p + \theta'_q q -\varphi_p p^2 - \varphi_q q^2 + \xi_{pq} pq)} \ket{\psi(0)}.
\end{equation}
In this equation, the phases generated by the self-Kerr terms are equal in modulus but opposite in sign to those in Eq.~(\ref{eq:evolvedStatePhi1Phi4}). We can therefore cancel out the effect of the self-Kerr terms by preparing the initial state $\ket{\psi(0)} = (\ket{\phi_1} + \ket{\phi_4})/\sqrt{2}$ and transforming it halfway through the time evolution into state $\ket{\psi'(t/2)} = (-\ket{\phi_2}+\ket{\phi_3})/\sqrt{2}$. {This  can be realised by the action of the spin operators $\sigma_1^x\sigma_2^z$ onto $\ket{\psi(0)}$}. At time $t$, the self-Kerr terms in the Hamiltonian vanish and, neglecting a global phase as before, we get a final state 
\begin{equation} \label{eq:finalStateECS}
\begin{split}
    \ket{\psi(t)} & = U(t/2) \sigma_1^x \sigma_2^z U(t/2) \ket{\psi(0)} \\ & \approx e^{-i t (\Theta_p p + \Theta_q q + \xi_{pq} pq)}\frac{(\ket{\phi_3} - \ket{\phi_2})}{\sqrt{2}}
\end{split}
\end{equation}
with $\Theta_p = {\epsilon_1 \epsilon_2 \eta_1^2 (\eta_2^2 - 2)}/({8 \lambda})$ and $\Theta_q$ obtained by taking $\eta_1\leftrightarrow\eta_2$ in $\Theta_p$. Tracing out the spin degrees of freedom, the resultant bosonic effective interaction Hamiltonian is
\begin{equation} \label{eq:noVarphi}
    H_\mathrm{ideal} = \Theta_p n_1 + \Theta_q n_2 + \xi_{pq} n_1n_2.
\end{equation}
{The cross-Kerr term can be further isolated} by applying the rotation operators $R_1 (\Theta_p) = e^{i \Theta_p n_1 t}$ and $R_2 (\Theta_q) = e^{i \Theta_q n_2 t}$ after the second time evolution operator. Note that the effective Hamiltonian including the cross-Kerr interaction is diagonal in the bosonic operators, as it stems from Eq.~\eqref{eq:hamBeforeMagnusCK}, and consequently it does not change the boson number of an initial state $\ket{p,q}$ containing $p$ and $q$ bosons for the first and second oscillator, respectively. Crucially, each Fock state $\ket{p,q}$ will acquire a different phase during the evolution dependent on $p$ and $q$ that can produce non-trivial entanglement dynamics among them, as we will show later.

\begin{figure*}[t]
\centering
\includegraphics[width=1\textwidth]{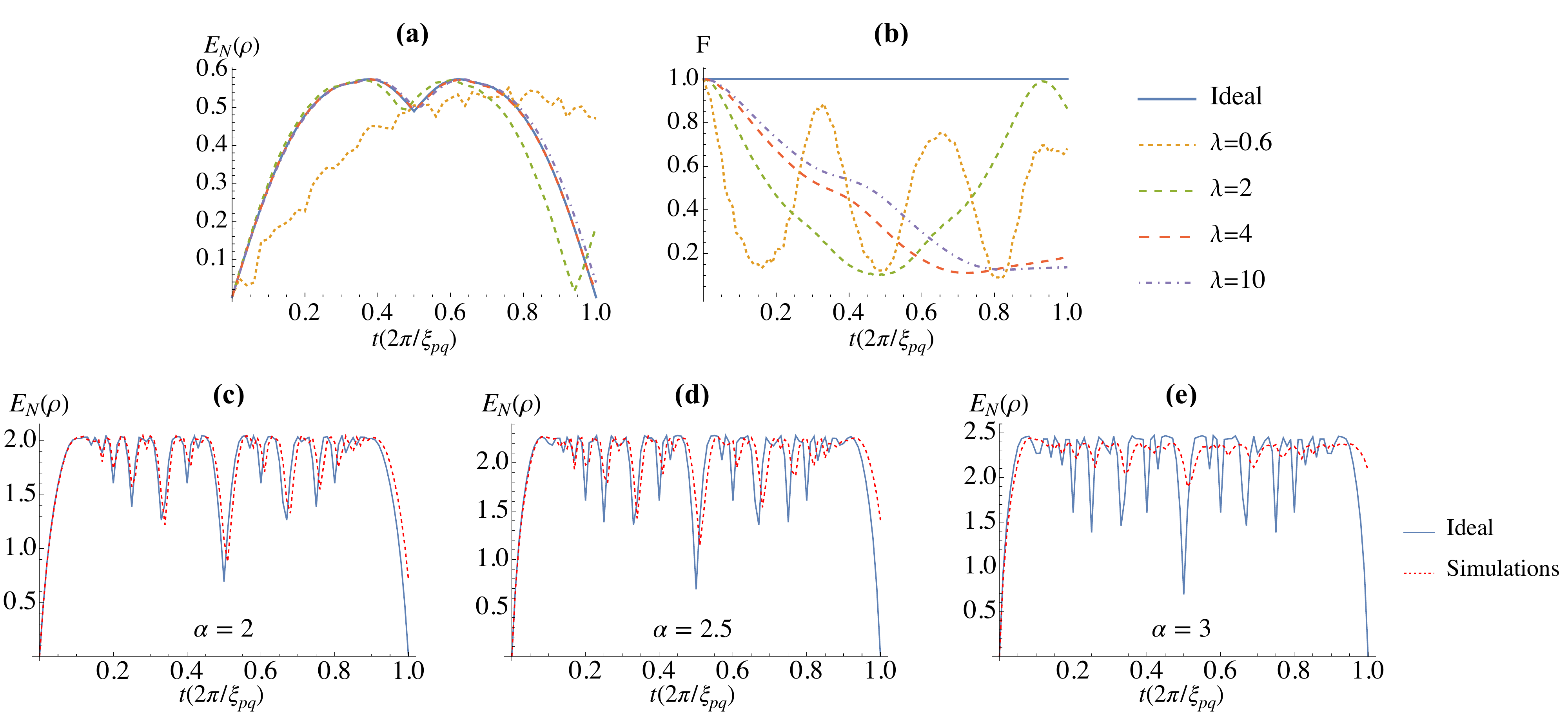}
\caption{{\bf (a)}-{\bf (b)}: Comparison of the state evolved through $H_\mathrm{linear}$ and the ideal state resulting from the effective Hamiltonian $H_\mathrm{ideal}$ in Eq.~(\ref{eq:noVarphi}). In these simulations we have used the parameters $\alpha_{1,2}= 0.5$, {$\epsilon_{1,2}/\omega=0.2, \eta_{1,2}/\omega=0.1$ with $\omega$ the value taken by the bosonic frequency (identical for both subsystems)}. In panel {\bf (a)} we report the logarithmic negativity, while the state fidelity is plotted in panel {\bf (b)}. {\bf (c)}-{\bf (e)}: Entanglement dynamics quantified with logarithmic negativity for increasing values of the coherent-state amplitudes. We have taken $\alpha_{1,2}=2$ in panel {\bf (c)}, $2.5$ in {\bf (d)} and, finally, $3$ in {\bf (e)}. The other parameter values are {$\epsilon_{1,2}/\omega=0.5, \eta_{1,2}/\omega=0.05$, and $\lambda/\omega=10$}.}
\label{fig:crossKerr}
\end{figure*}

It is worth emphasising that it is only possible to engineer the cross-Kerr Hamiltonian when the parameters take appropriate values allowing the necessary approximations to be made. Besides the set of requirements claimed throughout the derivation of our effective model, we need to ensure a large spin-spin coupling rate $\lambda$ {with respect to the parameters $\tilde{g}_{1,2}$}, so that $U_I (t) \approx \exp(-i t H_\mathrm{eff})$. {Such a restricted parameter regime constrains the strength of the achievable cross-Kerr coefficient $\xi_{pq}$ in Eq.~\eqref{eq:crossKerrCoeff}}, which in turn determines the amount of entanglement that could be generated between coherent states. There is thus a trade-off between the accuracy of the approximations made, the amount of entanglement generated between the coherent states, {and the total time of the evolution.}

\subsection{Results of the Simulations} \label{sec:resultsCrossKerr}

We compare the states evolved using the ideal cross-Kerr Hamiltonian and the original Hamiltonian model but without claim for any of the approximations stated above, that is
\begin{equation} \label{eq:linearHamCK}
    H_\mathrm{linear}=H^\mathrm{disp}_1 + H^\mathrm{disp}_2 + \lambda \sigma^z_1 \sigma^z_2
\end{equation} 
with $H^\mathrm{disp}_1 (H^\mathrm{disp}_2)$ as in Eq.~(\ref{eq:dispHam}) and acting on the first (second) spin-boson system. We first study the entanglement dynamics of the bosonic systems using logarithmic negativity~\cite{Plenio2005}. For a bipartite system comprising parties $A$ and $B$, this is given by
\begin{equation}
    E_N(\rho) = \log || \rho^{T_A} ||,
\end{equation}
where $\rho^{T_A}$ is the partial transpose of $\rho$ with respect to subsystem $A$ and $||\rho|| = \mathrm{Tr} \sqrt{\rho \rho^\dagger}$ is the trace norm. The results of our simulations are reported in Fig.~\ref{fig:crossKerr}{\bf (a)} 
for relatively small coherent-state amplitudes  $\alpha_{1,2}=0.5$ and confirm that we can replicate a cross-Kerr interaction using just linear interactions and spin drivings. However, as expected, the success of the protocol depends on the spin-spin coupling $\lambda$, as too small a value of this parameter results in significantly different entanglement dynamics.

In order to gauge the similarity of the states achieved through the two dynamical models, we now turn our attention to the state fidelity. For two generic density matrices $\rho_{A,B}$, this is defined as 
\begin{equation}
    F(\rho_A, \rho_B) = \left( \mathrm{Tr} \sqrt{\sqrt{\rho_A} \rho_B \sqrt{\rho_A}} \right)^2.
\end{equation}
{Remarkably, despite the very good agreement of the entanglement dynamics resulting from the two models being studied, Fig.~\ref{fig:crossKerr}{\bf (b)} shows that the ideal state is not retrieved by the effective evolution. Indeed, this is due to} local phases which are present in the simulated state but do not appear in the ideal one. Such phases stem from the Taylor expansion of the exponential in Eq.~\eqref{eq:sHam}
\begin{equation}
    e^{\eta (a(t)-a^\dagger(t))} = \sum_{n=0}^\infty \frac{\eta^n}{n!} (a(t)-a^\dagger(t))^n.
\end{equation}
{The expansion was truncated at $\eta^2$ in Eq.~\eqref{eq:noVarphi}. Yet, higher-order but diagonal terms may play a role (e.g. for $n=4$), which affects the dynamics}. This means that other elements enter into the effective Hamiltonian in Eq.~\eqref{eq:noVarphi}, including self-Kerr terms. These then generate the local phases that make the resulting state different from the ideal one. 

Evidence of the correctness of this argument is gathered already by considering the inclusion of the $n=4$ component in the ideal ECS and reconsidering the value taken by fidelity with the simulated state. For  {$\lambda/\omega=10$} and over the time period  $t/\xi_{pq}\in2\pi[0,5]$, values of state fidelity larger than 0.7 are achieved, thus demonstrating improved performances. Taking into account the $n=6$ component, however, does not result in yet higher fidelity; this improving effect is therefore not monotonic with the number of higher-order contributions included in the effective model. This could be a result of interference effects which emerge among successive higher-order terms. 

While ECS of small amplitudes are useful, we seek to entangle larger coherent states, which are more difficult to generate but play a more prominent role in protocols for quantum information processing. In Fig.~\ref{fig:crossKerr}{\bf (c)}-{\bf (e)}, we plot the entanglement dynamics of ideal ECS and simulated states using linear interactions for $\alpha_{1,2}=2, 2.5, 3$. We can appreciate a very good performance of the effective dynamics for $\alpha=2$, while the comparison with the target state worsens as $\alpha$ grows, as should be expected {for the considered parameters, in light of the Lamb-Dicke condition.} Nevertheless, though the dynamics vary significantly for $\alpha=3$, {a large amount of} entanglement is produced between the two states. 

The dips in the plots are due to the entangling phases applied to the Fock states $\ket{p,q}$ in Eq.~\eqref{eq:finalStateECS}. Were just one of these $\ket{p,q}$ elements to acquire a phase, the entanglement dynamics of the overall state would show entanglement falling to zero periodically with period dependent on the frequency of the phase. However, when the phases in Eq.~\eqref{eq:finalStateECS} are applied to all $\ket{p,q}$ elements in the coherent states, they all have different frequencies. Dips in the logarithmic negativity appear, then, at times depending on the periods of the phases of Fock state elements $\ket{p,q}$ with large contributions in the ECS. As higher $p$ and $q$ values result in higher frequencies, this effect is more pronounced for larger coherent state amplitudes $\alpha$ as we can see by comparing Figs.~\ref{fig:crossKerr} {\bf (a)} and {\bf (c)}-{\bf (e)}.

Surprisingly, the average entanglement $E_N^\mathrm{avg} (\rho)$ is larger for the simulated case than the ideal one, the difference between them growing with $\alpha$. Taking the average for $0.1 \leq t \leq 0.9$ so as to avoid the boundaries of the period (where entanglement rapidly rises and falls), for $\alpha=2$ we find that $E_N^\mathrm{avg} (\rho_\mathrm{sim}) = 1.869$ compared to $E_N^\mathrm{avg} (\rho_\mathrm{ideal}) = 1.863$, while for $\alpha = 2.5$ the average entanglement is $E_N^\mathrm{avg} (\rho_\mathrm{sim}) = 2.098$ and for the target state $E_N^\mathrm{avg} (\rho_\mathrm{ideal}) = 2.069$, and finally for $\alpha=3$ we have $E_N^\mathrm{avg} (\rho_\mathrm{sim}) = 2.304$ and $E_N^\mathrm{avg} (\rho_\mathrm{ideal}) = 2.233$. Despite the similarities in entanglement dynamics, the fidelities between the ideal and simulated states are again very poor. When we compare the state populations only, finding the fidelity between the states when off-diagonal terms are neglected, this increases dramatically. Optimising over time $t\in[0,1]$, the maximum infidelity $1-F$ is $2.912\times 10^{-7}$ for $\alpha = 2$, $2.016\times 10^{-5}$ for $\alpha=2.5$ and $7.279\times 10^{-4}$ for $\alpha=3$. The strong overlap between state populations is yet another indicator that the lack in fidelity between our target and simulated states is due to local phase factors.

\section{Engineering a Nonlinear Beam Splitter Interaction} \label{sec:NOON}

\subsection{Developing a Protocol} \label{sec:deriveNOON}

We now turn our attention to particular forms of N00M states, which generalize the well-known N00N family to  superpositions of asymmetric number states as follows 
\begin{equation} \label{eq:NOON}
    \ket{\mathrm{N00M}} = \frac{\ket{n,0}_{12}+i\ket{0,m}_{12}}{\sqrt{2}},
\end{equation}
with $\ket{p}_j$ a Fock state of the bosonic mode $j=1,2$ with $p$ excitations. In order to create states of this form, we would need to simulate a nonlinear beam splitter interaction. Starting again from Eqs.~\eqref{eq:HamAfterApprox} and \eqref{eq:totalHamiltonian}, in this case we take one spin rotation term with driving frequency $\Delta=-n\omega$ for the first spin system and $\Delta=-m\omega$ for the second (with $n,m$ as in Eq.~\eqref{eq:NOON}) and null phase for both. The Hamiltonian in each spin-boson system thus becomes a Jaynes-Cummings interaction, of the form 
\begin{equation} \label{eq:hamSpinBosonNOON}
    H_{\mathrm{SB}j} = k_j [\sigma^+_j a_j^{p_j} + \sigma^-_j (a_j^\dagger)^{p_j}],\qquad(j=1,2)
\end{equation}
where $k_j = \epsilon_j \eta_j^{p_j}/(2{p_j}!)$, $p_1=n$ and $p_2=m$. We now follow a similar approach as for the cross-Kerr case, passing through the move to the interaction picture with respect to the  spin-spin coupling $\lambda \sigma^z_1 \sigma^z_2$ and the Magnus expansion of the time-ordered evolution operator. As before, the latter reveals that most terms can be neglected. By using the basis elements 
\begin{equation} \label{eq:basisNLBS}
\begin{split}
    & \ket{\phi_1} = \ket{e,p}_1 \otimes \ket{e,q}_2, \\ 
    & \ket{\phi_2} = \ket{e,p}_1 \otimes \ket{g,q+m}_2, \\ 
    & \ket{\phi_3} = \ket{g,p+n}_1 \otimes \ket{e,q}_2, \\
    & \ket{\phi_4} = \ket{g,p+n}_1 \otimes \ket{g,q+m}_2,
\end{split}
\end{equation}
whose form is dictated by the conservation of the total number of excitations entailed by the non-linear beam-splitter dynamics, the effective Hamiltonian can again be written in matrix form as Eq.~(\ref{eq:matrixHam}) where now $\alpha = (\tilde{k}_1^2 + \tilde{k}_2^2)/(2\lambda)$ and $\beta = \tilde{k}_1 \tilde{k}_2 /\lambda$ with $\tilde{k}_1 = k_1 \sqrt{(p+n)!/p!}$ and $\tilde{k}_2 = k_2 \sqrt{(q+m)!/q!}$.

We see from Eq.~\eqref{eq:matrixHam} that acting on state $\ket{\phi_3}$ with $H_\mathrm{eff}$ will have the effect of making a superposition of $\ket{\phi_2}$ and $\ket{\phi_3}$. This is precisely what we need to generate N00M states. For the superposition to resemble the state in Eq.~\eqref{eq:NOON}, we need to take $p=q=0$. The initial state then takes the form $\ket{\phi_3} = \ket{g,n}_1 \otimes \ket{e,0}_2$. We also require a fixed evolution time $t=\pi/(4\beta)$ so that the two superposition elements have equal contribution in the state. The whole system of spins and bosons at time $t$ is
\begin{equation} \label{eq:idealNOON}
    \ket{\phi(t)} = \frac{e^{i\pi\alpha/(4\beta)}}{\sqrt{2}} \left( \ket{g,n}_1\otimes \ket{e,0}_2 + i \ket{e,0}_1 \otimes \ket{g,m}_2 \right).
\end{equation}

{To retrieve the bosonic N00M state and isolate it from the spins, we make use of a probabilistic protocol. That is, we measure the spins with either projectors $\Pi_{\pm x} = (\ket{e} \pm \ket{g})(\bra{e} \pm \bra{g})/2$ or $\Pi_{\pm y} = (\ket{e} \pm i \ket{g})(\bra{e} \mp i \bra{g})/2$ (ensuring that the same measurement is performed on each spin) before tracing them out of the system. The outcome of the measurement is irrelevant,  resulting in a N00M state of the same form as in Eq.~\eqref{eq:NOON}.} 

\subsection{Results of Simulations} \label{sec:resultsNOON}

\begin{figure*}[t]
\includegraphics[width=1\textwidth]{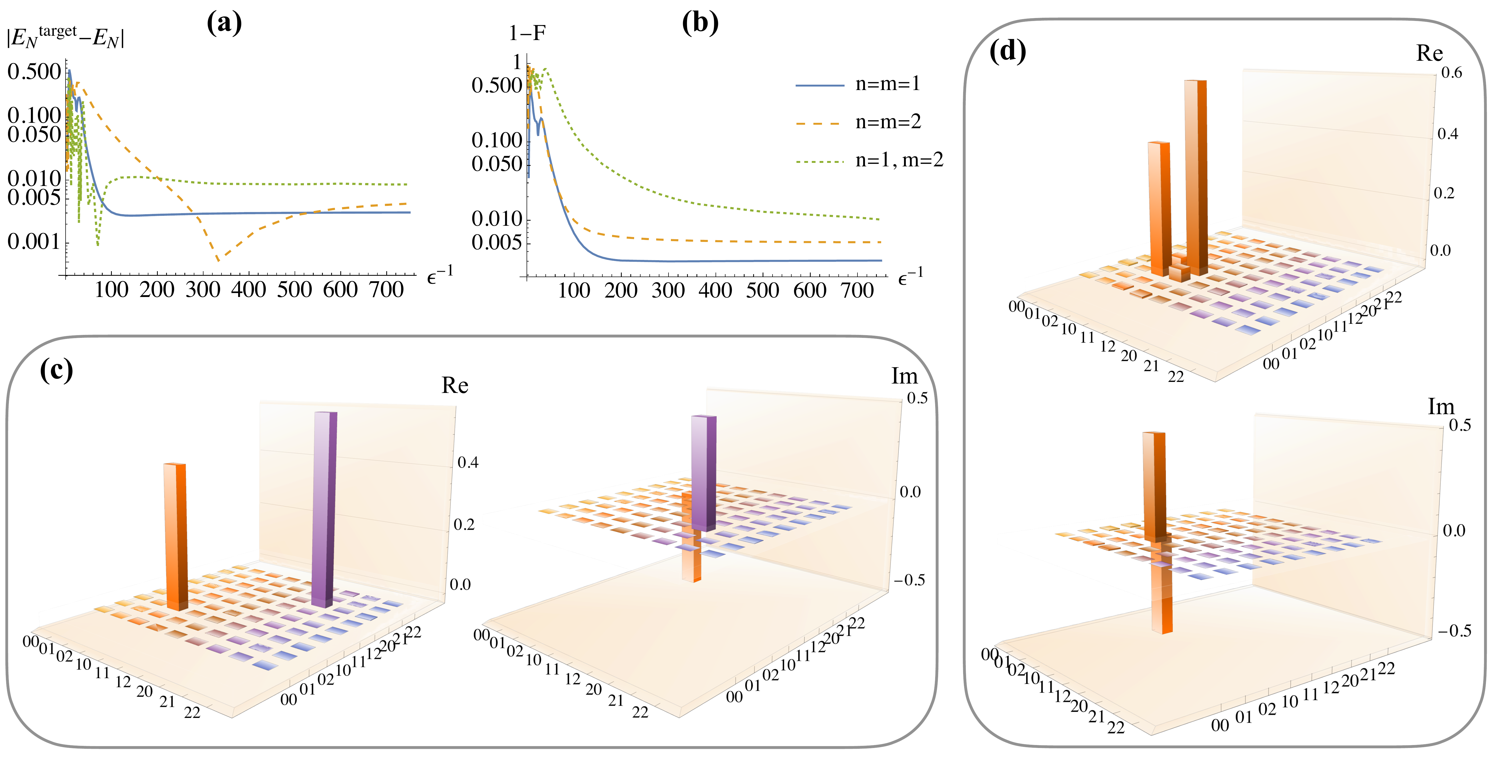}
\caption{{\bf (a)}: Difference between the logarithmic negativity of the target N00M state and the time evolved state. For $n=m=1$, we take {$\eta/\omega=0.25$ and $\epsilon = \lambda/6$. For $n=m=2$, we take $\eta/\omega=0.31$ and $\epsilon = 2\lambda/5$. For $n=1, m=2$, we take $\eta/\omega=0.35$ and $\epsilon = 5\lambda/34$.} {\bf (b)}: Infidelity between the same states as in {\bf (a)}. {\bf (c)}: Density matrix elements for $n=m=2$ {taking $\epsilon/\omega=1/750$.} 
{\bf (d)}: Density matrix elements for $n=1, m=2$ {taking $\epsilon/\omega=1/750$.}
}
\label{fig:NOON}
\end{figure*}
As done before for the cross-Kerr dynamics, we aim at characterizing the performance of the effective model by explicit comparison with the target state that we would like to accomplish. 
We begin from the Hamiltonian
\begin{equation} \label{eq:linearHamNLBS}
    H_\mathrm{linear} (t) = H_1^\mathrm{s} (t) + H_2^\mathrm{s} (t) + \lambda \sigma_1^z \sigma_2^z
\end{equation}
where $H_1^\mathrm{s} (t)$ $[H_2^\mathrm{s} (t)]$ is the spin-boson Hamiltonian in Eq.~(\ref{eq:sHam}) acting on the first [second] system. {It is worth recalling that, in this case, the Hamiltonian becomes time-dependent.}

{We evolve the system for the fixed time}
\begin{equation} \label{eq:timeParameter}
    t= \frac{\pi}{4\beta} = \frac{\pi \lambda \sqrt{n!m!}}{\epsilon_1 \epsilon_2 \eta_1^n \eta_2^m},
\end{equation}
so that the time needed to produce N00M states with increasing $n$ and $m$ quickly grows. We aim to limit the length of time needed to achieve such states, in particular by choosing small values of $\lambda$ and large values of $\epsilon$ and $\eta$. However, each of the above conditions directly contradicts the limitations we put in place for the approximations to be accurate. Therefore, as in the cross-Kerr case, there is a trade off and we must find a compromise between restricting the needed evolution time and maintaining a high accuracy of the approximations. We can use the time-independent Hamiltonian in Eqs.~\eqref{eq:totalHamiltonian} and \eqref{eq:hamSpinBosonNOON} to find a suitable relations between the parameters; when the value of $\eta$ (in units of $\omega)$ is fixed, the time-evolved state is determined only by the ratio between $\epsilon$ and $\lambda$. A close-to-unit fidelity of the time-evolved state with the ideal N00M state shows that the approximation we made using the Magnus expansion is highly accurate for our chosen parameter regime.

We show the results of our simulations in Fig.~\ref{fig:NOON}. We generate three states: the first is a Bell-like N00N state with $n=m=1$, the second has $n=m=2$, while  the third one is a N00M state where $n=1$ and $m=2$. The difference $\vert E^\text{target}_N-E_N\vert$ between the logarithmic negativity of the target states and the time-evolved ones is shown in Fig.~\ref{fig:NOON}{\bf(a)}. Clearly, we  need {small values of $\epsilon$ and $\lambda$} for the approximations in our analysis to be valid. The achievement of such a regime entails promising results. For the $n=m=2$ state, we need especially {low} values; the entanglement does not start to level off before around {$\epsilon/\omega=1/500$}, and the difference with its target settles at a value of around $4\times10^{-3}$. {However, this is not the} minimum in entanglement difference that we can achieve, which instead occurs at $\epsilon/\omega \approx 1/340$. This is because the logarithmic negativity of the simulated state is much higher than that of the ideal N00N state 
{when $\epsilon$ is relatively large,} but then falls to be slightly lower than the target before it converges. 

\begin{figure}[b]
\centering
\includegraphics[width=1\columnwidth]{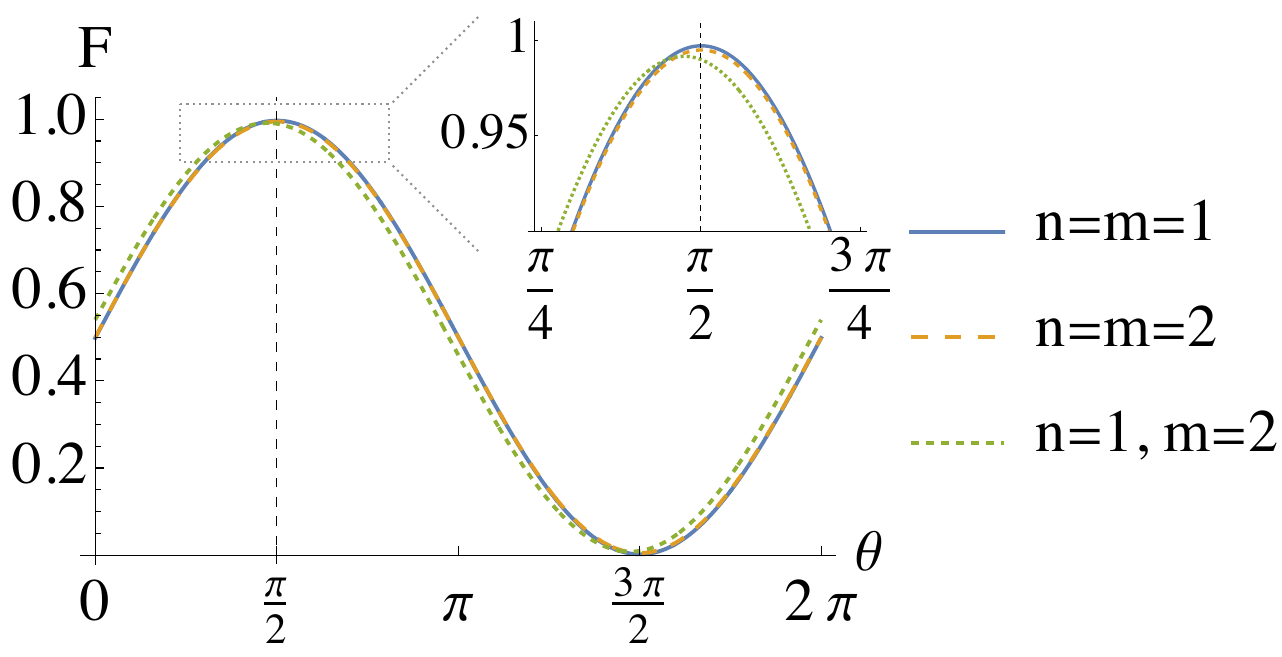}
\caption{Fidelity between the state evolved with $H_\mathrm{linear} (t)$ and the N00N state $\ket{n,0}+e^{i\theta}\ket{0,m}$. For each case, the driving strength is taken to be $\epsilon=1/750$.}
\label{fig:GenNOONfidelity}
\end{figure}

The {infidelity} $1-F$ of each simulated state with its corresponding ideal one is shown in Fig.~\ref{fig:NOON}{\bf (b)}. We achieve a much better result than for the ECS, {due to the absence of different local phases among many populated Fock states, as is the case for the ECS}. It is clear that increasing the number of excitations $n$ or $m$ limits the success of the protocol {due, once more, to the constraint set by the Lamb-Dicke regime}. 

We also note here that, though we select specific values for $\eta$ and $\epsilon$ in Fig.~\ref{fig:NOON} using Eqs.~\eqref{eq:totalHamiltonian} and \eqref{eq:hamSpinBosonNOON} as explained above, we obtain similar results when they are varied. For instance, we checked the reliability of our results for $n=m=1$ by obtaining results for $0.2\leq \eta \leq 0.3$ and $0.05\leq \epsilon/\lambda \leq 0.25$. The corresponding range of fidelities is $\{0.99411,0.99856\}$, compared to our original value of 0.99694, and entanglement difference goes between $\{0.001164,0.005663\}$ with an original value of 0.003053. Fidelity and entanglement are both broadly increasing as $\eta$ decreases. We carried out the same analysis for the parameters in the $n=m=2$ and $n=1$, $m=2$ cases and achieved comparable results.

To get a deeper insight into the similarities between target and actual states, we study the density matrices of the bosonic modes, whose real and imaginary elements are plotted in Fig.~\ref{fig:NOON}{\bf (c)} for $n=m=2$ and {\bf (d)} for $n=1,m=2$. We take $\epsilon/\omega=1/750$, a value for which the resemblance to the desired N00N state is evident. However, in both cases we lack the symmetry we should see, especially in the plots of the real parts of the density matrix elements. Half of the population of $\ket{n,0}$ should have been transferred to state $\ket{0,m}$ through the course of the time evolution but this did not occur perfectly. For instance, for $n=m=2$, the final population of $\ket{0,n}$ was 0.574 instead of 0.5.

In order to bypass potential limitations in the performance of our comparisons, we next consider a class of N00M states having the form
\begin{equation}
    \ket{\mathrm{N00M}_\mathrm{gen}} = \frac{\ket{n,0}+e^{i\theta} \ket{0,m}}{\sqrt{2}}
\end{equation}
with $\theta$ a phase  whose value we change to look for potential better comparators with the achieved state. 
The fidelities with varying $\theta$ are shown in Fig.~\ref{fig:GenNOONfidelity}. As expected, we find that the maximum fidelity for each case is reached for values of $\theta$ close to $\pi/2$, corresponding to the ideal N00N state. For $n=m=1$, 
the difference is negligible: fidelity is maximised for $\theta\gtrsim\pi/2$ and the increase in fidelity is in the order of $10^{-6}$. {There is no difference at all for $n=m=2$, where the fidelity is maximised for $\pi/2$ exactly.} However, there is a more noticeable gain for the N00M state as {the maximum occurs slightly away from $\theta=\pi/2$, leading to an increase of $10^{-3}$ in the fidelity.}

\section{Conclusions} \label{sec:conc}
We have proposed two protocols for generating ECS and N00M states of two bosons requiring only linear spin-spin and spin-boson interactions and spin rotations. Such effective interaction mechanisms hold under a suitable parameter regime, namely, strong coupling between the spins, small spin driving amplitudes and large bosonic frequencies. However, the above conditions also come with restrictions. In the case of replicating a cross-Kerr Hamiltonian, they limit the amount of entanglement that can be generated, while also for N00N states, they escalate the total time evolution needed. Suitable parameters can nevertheless be found so that the requirements are fulfilled and we can successfully replicate the entanglement dynamics of an ECS for coherent state amplitudes of $\alpha>2$ and simulate N00M states with high fidelity and matching entanglement values. 

The coherent state amplitudes achieved in our work go beyond what has currently been experimentally realised~\cite{Ourjoumtsev2009,Israel2019,Wang2016,Jeon2024}, as this method can generate ECSs with amplitudes over 1.92~\cite{Wang2016}. We also generate considerably higher entanglement than other proposals; as Ref.~\cite{Kounalakis2023} proposes a method to create ECSs proportional to $\ket{0,0} + \ket{\alpha, \alpha}$, they are restricted to a maximum of $E_N = \ln 2$ under ideal conditions, even when $\alpha$ is as large as 2. Meanwhile, we can reach $E_N=2$ for the same amplitude. Moreover, our scheme has the added benefit of flexibility with regards to physical systems we can use to implement it.

While we see more limitations in the size of the N00N states we can engineer compared to other works~\cite{Jones2009,Zhang2018}, our protocol brings the advantage of the ability to create N00M states. We can achieve a fidelity of 0.989 with our protocol for the state $\ket{10}+\ket{02}$, where Ref.~\cite{Wang2011} can reach only $F=0.42$. We can also obtain high-fidelity N00N states for $n=m=2$; we get a fidelity of 0.994. While higher fidelities can be obtained for optical N00N states~\cite{Afek2010}, our protocol performs well against other proposals limited to fidelities of 0.861~\cite{Su2014}, 0.78~\cite{Grebel2024} and 0.5~\cite{Wang2011}. We hope to follow this work with a study of the effects of decoherence and losses on the performance of the protocol in due time.

\acknowledgements

This work is supported by the H2020-FETOPEN-2018-2020 project TEQ (grant No. 766900), the Horizon Europe EIC Pathfinder project  QuCoM (Grant Agreement No.\,101046973), the Royal Society Wolfson Research Fellowship (RSWF$\backslash$R3$\backslash$183013), the Royal Society International Exchanges Programme (IEC$\backslash$R2$\backslash$192220), the Leverhulme Trust Research Project Grant (grant No. RGP-2018-266), the UK EPSRC (grant EP/T028424/1), the Department for the Economy Northern Ireland under the US-Ireland R\&D Partnership Programme, and by the Ram{\'o}n y Cajal (RYC2023-044095-I) research fellowship. 

\bibliography{bosonBib.bib}

\begin{thebibliography}{41}%
\makeatletter
\providecommand \@ifxundefined [1]{%
 \@ifx{#1\undefined}
}%
\providecommand \@ifnum [1]{%
 \ifnum #1\expandafter \@firstoftwo
 \else \expandafter \@secondoftwo
 \fi
}%
\providecommand \@ifx [1]{%
 \ifx #1\expandafter \@firstoftwo
 \else \expandafter \@secondoftwo
 \fi
}%
\providecommand \natexlab [1]{#1}%
\providecommand \enquote  [1]{``#1''}%
\providecommand \bibnamefont  [1]{#1}%
\providecommand \bibfnamefont [1]{#1}%
\providecommand \citenamefont [1]{#1}%
\providecommand \href@noop [0]{\@secondoftwo}%
\providecommand \href [0]{\begingroup \@sanitize@url \@href}%
\providecommand \@href[1]{\@@startlink{#1}\@@href}%
\providecommand \@@href[1]{\endgroup#1\@@endlink}%
\providecommand \@sanitize@url [0]{\catcode `\\12\catcode `\$12\catcode
  `\&12\catcode `\#12\catcode `\^12\catcode `\_12\catcode `\%12\relax}%
\providecommand \@@startlink[1]{}%
\providecommand \@@endlink[0]{}%
\providecommand \url  [0]{\begingroup\@sanitize@url \@url }%
\providecommand \@url [1]{\endgroup\@href {#1}{\urlprefix }}%
\providecommand \urlprefix  [0]{URL }%
\providecommand \Eprint [0]{\href }%
\providecommand \doibase [0]{https://doi.org/}%
\providecommand \selectlanguage [0]{\@gobble}%
\providecommand \bibinfo  [0]{\@secondoftwo}%
\providecommand \bibfield  [0]{\@secondoftwo}%
\providecommand \translation [1]{[#1]}%
\providecommand \BibitemOpen [0]{}%
\providecommand \bibitemStop [0]{}%
\providecommand \bibitemNoStop [0]{.\EOS\space}%
\providecommand \EOS [0]{\spacefactor3000\relax}%
\providecommand \BibitemShut  [1]{\csname bibitem#1\endcsname}%
\let\auto@bib@innerbib\@empty
\bibitem [{\citenamefont {Sanders}(2012)}]{Sanders2012}%
  \BibitemOpen
  \bibfield  {author} {\bibinfo {author} {\bibfnamefont {B.~C.}\ \bibnamefont
  {Sanders}},\ }\bibfield  {title} {\bibinfo {title} {Review of entangled
  coherent states},\ }\href
  {https://doi.org/https://doi.org/10.1088/1751-8113/45/24/244002} {\bibfield
  {journal} {\bibinfo  {journal} {J. Phys. A: Math. Theor.}\ }\textbf {\bibinfo
  {volume} {45}},\ \bibinfo {pages} {244002} (\bibinfo {year}
  {2012})}\BibitemShut {NoStop}%
\bibitem [{\citenamefont {Sangouard}\ \emph {et~al.}(2010)\citenamefont
  {Sangouard}, \citenamefont {Simon}, \citenamefont {Gisin}, \citenamefont
  {Laurat}, \citenamefont {Tualle-Brouri},\ and\ \citenamefont
  {Grangier}}]{Sangouard2010}%
  \BibitemOpen
  \bibfield  {author} {\bibinfo {author} {\bibfnamefont {N.}~\bibnamefont
  {Sangouard}}, \bibinfo {author} {\bibfnamefont {C.}~\bibnamefont {Simon}},
  \bibinfo {author} {\bibfnamefont {N.}~\bibnamefont {Gisin}}, \bibinfo
  {author} {\bibfnamefont {J.}~\bibnamefont {Laurat}}, \bibinfo {author}
  {\bibfnamefont {R.}~\bibnamefont {Tualle-Brouri}},\ and\ \bibinfo {author}
  {\bibfnamefont {P.}~\bibnamefont {Grangier}},\ }\bibfield  {title} {\bibinfo
  {title} {Quantum repeaters with entangled coherent states},\ }\href
  {https://doi.org/https://doi.org/10.1364/JOSAB.27.00A137} {\bibfield
  {journal} {\bibinfo  {journal} {J. Opt. Soc. Am. B}\ }\textbf {\bibinfo
  {volume} {27}},\ \bibinfo {pages} {A137} (\bibinfo {year}
  {2010})}\BibitemShut {NoStop}%
\bibitem [{\citenamefont {Brask}\ \emph {et~al.}(2010)\citenamefont {Brask},
  \citenamefont {Rigas}, \citenamefont {Polzik}, \citenamefont {Andersen},\
  and\ \citenamefont {S\o{}rensen}}]{Brask2010}%
  \BibitemOpen
  \bibfield  {author} {\bibinfo {author} {\bibfnamefont {J.~B.}\ \bibnamefont
  {Brask}}, \bibinfo {author} {\bibfnamefont {I.}~\bibnamefont {Rigas}},
  \bibinfo {author} {\bibfnamefont {E.~S.}\ \bibnamefont {Polzik}}, \bibinfo
  {author} {\bibfnamefont {U.~L.}\ \bibnamefont {Andersen}},\ and\ \bibinfo
  {author} {\bibfnamefont {A.~S.}\ \bibnamefont {S\o{}rensen}},\ }\bibfield
  {title} {\bibinfo {title} {Hybrid long-distance entanglement distribution
  protocol},\ }\href
  {https://doi.org/https://doi.org/10.1103/PhysRevLett.105.160501} {\bibfield
  {journal} {\bibinfo  {journal} {Phys. Rev. Lett.}\ }\textbf {\bibinfo
  {volume} {105}},\ \bibinfo {pages} {160501} (\bibinfo {year}
  {2010})}\BibitemShut {NoStop}%
\bibitem [{\citenamefont {Lim}\ \emph {et~al.}(2016)\citenamefont {Lim},
  \citenamefont {Joo}, \citenamefont {Spiller},\ and\ \citenamefont
  {Jeong}}]{Lim2016}%
  \BibitemOpen
  \bibfield  {author} {\bibinfo {author} {\bibfnamefont {Y.}~\bibnamefont
  {Lim}}, \bibinfo {author} {\bibfnamefont {J.}~\bibnamefont {Joo}}, \bibinfo
  {author} {\bibfnamefont {T.~P.}\ \bibnamefont {Spiller}},\ and\ \bibinfo
  {author} {\bibfnamefont {H.}~\bibnamefont {Jeong}},\ }\bibfield  {title}
  {\bibinfo {title} {Loss-resilient photonic entanglement swapping using
  optical hybrid states},\ }\href
  {https://doi.org/https://doi.org/10.1103/PhysRevA.94.062337} {\bibfield
  {journal} {\bibinfo  {journal} {Phys. Rev. A}\ }\textbf {\bibinfo {volume}
  {94}},\ \bibinfo {pages} {062337} (\bibinfo {year} {2016})}\BibitemShut
  {NoStop}%
\bibitem [{\citenamefont {{van Enk}}\ and\ \citenamefont
  {Hirota}(2001)}]{vanEnk2001}%
  \BibitemOpen
  \bibfield  {author} {\bibinfo {author} {\bibfnamefont {S.~J.}\ \bibnamefont
  {{van Enk}}}\ and\ \bibinfo {author} {\bibfnamefont {O.}~\bibnamefont
  {Hirota}},\ }\bibfield  {title} {\bibinfo {title} {Entangled coherent states:
  Teleportation and decoherence},\ }\href
  {https://doi.org/https://doi.org/10.1103/PhysRevA.64.022313} {\bibfield
  {journal} {\bibinfo  {journal} {Phys. Rev. A}\ }\textbf {\bibinfo {volume}
  {64}},\ \bibinfo {pages} {022313} (\bibinfo {year} {2001})}\BibitemShut
  {NoStop}%
\bibitem [{\citenamefont {Wang}(2001)}]{Wang2001}%
  \BibitemOpen
  \bibfield  {author} {\bibinfo {author} {\bibfnamefont {X.}~\bibnamefont
  {Wang}},\ }\bibfield  {title} {\bibinfo {title} {Quantum teleportation of
  entangled coherent states},\ }\href
  {https://doi.org/https://doi.org/10.1103/PhysRevA.64.022302} {\bibfield
  {journal} {\bibinfo  {journal} {Phys. Rev. A}\ }\textbf {\bibinfo {volume}
  {64}},\ \bibinfo {pages} {022302} (\bibinfo {year} {2001})}\BibitemShut
  {NoStop}%
\bibitem [{\citenamefont {Jeong}\ \emph {et~al.}(2001)\citenamefont {Jeong},
  \citenamefont {Kim},\ and\ \citenamefont {Lee}}]{Jeong2001}%
  \BibitemOpen
  \bibfield  {author} {\bibinfo {author} {\bibfnamefont {H.}~\bibnamefont
  {Jeong}}, \bibinfo {author} {\bibfnamefont {M.~S.}\ \bibnamefont {Kim}},\
  and\ \bibinfo {author} {\bibfnamefont {J.}~\bibnamefont {Lee}},\ }\bibfield
  {title} {\bibinfo {title} {Quantum-information processing for a coherent
  superposition state via a mixed entangled coherent channel},\ }\href
  {https://doi.org/https://doi.org/10.1103/PhysRevA.64.052308} {\bibfield
  {journal} {\bibinfo  {journal} {Phys. Rev. A}\ }\textbf {\bibinfo {volume}
  {64}},\ \bibinfo {pages} {052308} (\bibinfo {year} {2001})}\BibitemShut
  {NoStop}%
\bibitem [{\citenamefont {Joo}\ and\ \citenamefont {Ginossar}(2016)}]{Joo2016}%
  \BibitemOpen
  \bibfield  {author} {\bibinfo {author} {\bibfnamefont {J.}~\bibnamefont
  {Joo}}\ and\ \bibinfo {author} {\bibfnamefont {E.}~\bibnamefont {Ginossar}},\
  }\bibfield  {title} {\bibinfo {title} {Efficient scheme for hybrid
  teleportation via entangled coherent states in circuit quantum
  electrodynamics},\ }\href {https://doi.org/https://doi.org/10.1038/srep26338}
  {\bibfield  {journal} {\bibinfo  {journal} {Sci. Rep.}\ }\textbf {\bibinfo
  {volume} {6}},\ \bibinfo {pages} {26338} (\bibinfo {year}
  {2016})}\BibitemShut {NoStop}%
\bibitem [{\citenamefont {Jeong}\ and\ \citenamefont {Kim}(2002)}]{Jeong2002}%
  \BibitemOpen
  \bibfield  {author} {\bibinfo {author} {\bibfnamefont {H.}~\bibnamefont
  {Jeong}}\ and\ \bibinfo {author} {\bibfnamefont {M.~S.}\ \bibnamefont
  {Kim}},\ }\bibfield  {title} {\bibinfo {title} {Efficient quantum computation
  using coherent states},\ }\href
  {https://doi.org/https://doi.org/10.1103/PhysRevA.65.042305} {\bibfield
  {journal} {\bibinfo  {journal} {Phys. Rev. A}\ }\textbf {\bibinfo {volume}
  {65}},\ \bibinfo {pages} {042305} (\bibinfo {year} {2002})}\BibitemShut
  {NoStop}%
\bibitem [{\citenamefont {Boto}\ \emph {et~al.}(2000)\citenamefont {Boto},
  \citenamefont {Kok}, \citenamefont {Abrams}, \citenamefont {Braunstein},
  \citenamefont {Williams},\ and\ \citenamefont {Dowling}}]{Boto2000}%
  \BibitemOpen
  \bibfield  {author} {\bibinfo {author} {\bibfnamefont {A.~N.}\ \bibnamefont
  {Boto}}, \bibinfo {author} {\bibfnamefont {P.}~\bibnamefont {Kok}}, \bibinfo
  {author} {\bibfnamefont {D.~S.}\ \bibnamefont {Abrams}}, \bibinfo {author}
  {\bibfnamefont {S.~L.}\ \bibnamefont {Braunstein}}, \bibinfo {author}
  {\bibfnamefont {C.~P.}\ \bibnamefont {Williams}},\ and\ \bibinfo {author}
  {\bibfnamefont {J.~P.}\ \bibnamefont {Dowling}},\ }\bibfield  {title}
  {\bibinfo {title} {Quantum interferometric optical lithography: exploiting
  entanglement to beat the diffraction limit},\ }\href
  {https://doi.org/https://doi.org/10.1103/PhysRevLett.85.2733} {\bibfield
  {journal} {\bibinfo  {journal} {Phys. Rev. Lett.}\ }\textbf {\bibinfo
  {volume} {85}},\ \bibinfo {pages} {2733} (\bibinfo {year}
  {2000})}\BibitemShut {NoStop}%
\bibitem [{\citenamefont {{D'Angelo}}\ \emph {et~al.}(2001)\citenamefont
  {{D'Angelo}}, \citenamefont {Chekhova},\ and\ \citenamefont
  {Shih}}]{DAngelo2001}%
  \BibitemOpen
  \bibfield  {author} {\bibinfo {author} {\bibfnamefont {M.}~\bibnamefont
  {{D'Angelo}}}, \bibinfo {author} {\bibfnamefont {M.~V.}\ \bibnamefont
  {Chekhova}},\ and\ \bibinfo {author} {\bibfnamefont {Y.}~\bibnamefont
  {Shih}},\ }\bibfield  {title} {\bibinfo {title} {Two-photon diffraction and
  quantum lithography},\ }\href
  {https://doi.org/https://doi.org/10.1103/PhysRevLett.87.013602} {\bibfield
  {journal} {\bibinfo  {journal} {Phys. Rev. Lett.}\ }\textbf {\bibinfo
  {volume} {87}},\ \bibinfo {pages} {013602} (\bibinfo {year}
  {2001})}\BibitemShut {NoStop}%
\bibitem [{\citenamefont {Kok}\ \emph {et~al.}(2004)\citenamefont {Kok},
  \citenamefont {Braunstein},\ and\ \citenamefont {Dowling}}]{Kok2004}%
  \BibitemOpen
  \bibfield  {author} {\bibinfo {author} {\bibfnamefont {P.}~\bibnamefont
  {Kok}}, \bibinfo {author} {\bibfnamefont {S.~L.}\ \bibnamefont
  {Braunstein}},\ and\ \bibinfo {author} {\bibfnamefont {J.~P.}\ \bibnamefont
  {Dowling}},\ }\bibfield  {title} {\bibinfo {title} {Quantum lithography,
  entanglement and {Heisenberg}-limited parameter estimation},\ }\href
  {https://doi.org/https://doi.org/10.1088/1464-4266/6/8/029} {\bibfield
  {journal} {\bibinfo  {journal} {J. Opt. B: Quantum Semiclass. Opt.}\ }\textbf
  {\bibinfo {volume} {6}},\ \bibinfo {pages} {S811} (\bibinfo {year}
  {2004})}\BibitemShut {NoStop}%
\bibitem [{\citenamefont {Nagata}\ \emph {et~al.}(2007)\citenamefont {Nagata},
  \citenamefont {Okamoto}, \citenamefont {{O'Brien}}, \citenamefont {Sasaki},\
  and\ \citenamefont {Takeuchi}}]{Nagata2007}%
  \BibitemOpen
  \bibfield  {author} {\bibinfo {author} {\bibfnamefont {T.}~\bibnamefont
  {Nagata}}, \bibinfo {author} {\bibfnamefont {R.}~\bibnamefont {Okamoto}},
  \bibinfo {author} {\bibfnamefont {J.~L.}\ \bibnamefont {{O'Brien}}}, \bibinfo
  {author} {\bibfnamefont {K.}~\bibnamefont {Sasaki}},\ and\ \bibinfo {author}
  {\bibfnamefont {S.}~\bibnamefont {Takeuchi}},\ }\bibfield  {title} {\bibinfo
  {title} {Beating the standard quantum limit with four-entangled photons},\
  }\href {https://doi.org/https://doi.org/10.1126/science.1138007} {\bibfield
  {journal} {\bibinfo  {journal} {Science}\ }\textbf {\bibinfo {volume}
  {316}},\ \bibinfo {pages} {726} (\bibinfo {year} {2007})}\BibitemShut
  {NoStop}%
\bibitem [{\citenamefont {Dowling}(2008)}]{Dowling2008}%
  \BibitemOpen
  \bibfield  {author} {\bibinfo {author} {\bibfnamefont {J.~P.}\ \bibnamefont
  {Dowling}},\ }\bibfield  {title} {\bibinfo {title} {Quantum optical metrology
  - the lowdown on high-{N00N} states},\ }\href
  {https://doi.org/https://doi.org/10.1080/00107510802091298} {\bibfield
  {journal} {\bibinfo  {journal} {Contemp. Phys.}\ }\textbf {\bibinfo {volume}
  {49}},\ \bibinfo {pages} {125} (\bibinfo {year} {2008})}\BibitemShut
  {NoStop}%
\bibitem [{\citenamefont {Israel}\ \emph {et~al.}(2014)\citenamefont {Israel},
  \citenamefont {Rosen},\ and\ \citenamefont {Silberberg}}]{Israel2014}%
  \BibitemOpen
  \bibfield  {author} {\bibinfo {author} {\bibfnamefont {Y.}~\bibnamefont
  {Israel}}, \bibinfo {author} {\bibfnamefont {S.}~\bibnamefont {Rosen}},\ and\
  \bibinfo {author} {\bibfnamefont {Y.}~\bibnamefont {Silberberg}},\ }\bibfield
   {title} {\bibinfo {title} {Supersensitive polarization microscopy using noon
  states of light},\ }\href
  {https://doi.org/https://doi.org/10.1103/PhysRevLett.112.103604} {\bibfield
  {journal} {\bibinfo  {journal} {Phys. Rev. Lett.}\ }\textbf {\bibinfo
  {volume} {112}},\ \bibinfo {pages} {103604} (\bibinfo {year}
  {2014})}\BibitemShut {NoStop}%
\bibitem [{\citenamefont {Slussarenko}\ \emph {et~al.}(2017)\citenamefont
  {Slussarenko}, \citenamefont {Weston}, \citenamefont {Chrzanowski},
  \citenamefont {Shalm}, \citenamefont {Verma}, \citenamefont {Nam},\ and\
  \citenamefont {Pryde}}]{Slussarenko2017}%
  \BibitemOpen
  \bibfield  {author} {\bibinfo {author} {\bibfnamefont {S.}~\bibnamefont
  {Slussarenko}}, \bibinfo {author} {\bibfnamefont {M.~M.}\ \bibnamefont
  {Weston}}, \bibinfo {author} {\bibfnamefont {H.~M.}\ \bibnamefont
  {Chrzanowski}}, \bibinfo {author} {\bibfnamefont {L.~K.}\ \bibnamefont
  {Shalm}}, \bibinfo {author} {\bibfnamefont {V.~B.}\ \bibnamefont {Verma}},
  \bibinfo {author} {\bibfnamefont {S.~W.}\ \bibnamefont {Nam}},\ and\ \bibinfo
  {author} {\bibfnamefont {G.~J.}\ \bibnamefont {Pryde}},\ }\bibfield  {title}
  {\bibinfo {title} {Unconditional violation of the shot-noise limit in
  photonic quantum metrology},\ }\href
  {https://doi.org/https://doi.org/10.1038/s41566-017-0011-5} {\bibfield
  {journal} {\bibinfo  {journal} {Nat. Photon.}\ }\textbf {\bibinfo {volume}
  {11}},\ \bibinfo {pages} {700} (\bibinfo {year} {2017})}\BibitemShut
  {NoStop}%
\bibitem [{\citenamefont {Casanova}\ \emph {et~al.}(2018)\citenamefont
  {Casanova}, \citenamefont {Puebla}, \citenamefont {Moya-Cessa},\ and\
  \citenamefont {Plenio}}]{Casanova2018}%
  \BibitemOpen
  \bibfield  {author} {\bibinfo {author} {\bibfnamefont {J.}~\bibnamefont
  {Casanova}}, \bibinfo {author} {\bibfnamefont {R.}~\bibnamefont {Puebla}},
  \bibinfo {author} {\bibfnamefont {H.}~\bibnamefont {Moya-Cessa}},\ and\
  \bibinfo {author} {\bibfnamefont {M.~B.}\ \bibnamefont {Plenio}},\ }\bibfield
   {title} {\bibinfo {title} {Connecting nth order generalised quantum {Rabi}
  models: Emergence of nonlinear spin-boson coupling via spin rotations},\
  }\href {https://doi.org/https://doi.org/10.1038/s41534-018-0096-9} {\bibfield
   {journal} {\bibinfo  {journal} {npj Quantum Inf.}\ }\textbf {\bibinfo
  {volume} {4}},\ \bibinfo {pages} {47} (\bibinfo {year} {2018})}\BibitemShut
  {NoStop}%
\bibitem [{\citenamefont {Ourjoumtsev}\ \emph {et~al.}(2009)\citenamefont
  {Ourjoumtsev}, \citenamefont {Ferreyrol}, \citenamefont {Tualle-Brouri},\
  and\ \citenamefont {Grangier}}]{Ourjoumtsev2009}%
  \BibitemOpen
  \bibfield  {author} {\bibinfo {author} {\bibfnamefont {A.}~\bibnamefont
  {Ourjoumtsev}}, \bibinfo {author} {\bibfnamefont {F.}~\bibnamefont
  {Ferreyrol}}, \bibinfo {author} {\bibfnamefont {R.}~\bibnamefont
  {Tualle-Brouri}},\ and\ \bibinfo {author} {\bibfnamefont {P.}~\bibnamefont
  {Grangier}},\ }\bibfield  {title} {\bibinfo {title} {Preparation of non-local
  superpositions of quasi-classical light states},\ }\href
  {https://doi.org/https://doi.org/10.1038/nphys1199} {\bibfield  {journal}
  {\bibinfo  {journal} {Nat. Phys.}\ }\textbf {\bibinfo {volume} {5}},\
  \bibinfo {pages} {189} (\bibinfo {year} {2009})}\BibitemShut {NoStop}%
\bibitem [{\citenamefont {Israel}\ \emph {et~al.}(2019)\citenamefont {Israel},
  \citenamefont {Cohen}, \citenamefont {Song}, \citenamefont {Joo},
  \citenamefont {Eisenberg},\ and\ \citenamefont {Silberberg}}]{Israel2019}%
  \BibitemOpen
  \bibfield  {author} {\bibinfo {author} {\bibfnamefont {Y.}~\bibnamefont
  {Israel}}, \bibinfo {author} {\bibfnamefont {L.}~\bibnamefont {Cohen}},
  \bibinfo {author} {\bibfnamefont {X.-B.}\ \bibnamefont {Song}}, \bibinfo
  {author} {\bibfnamefont {J.}~\bibnamefont {Joo}}, \bibinfo {author}
  {\bibfnamefont {H.~S.}\ \bibnamefont {Eisenberg}},\ and\ \bibinfo {author}
  {\bibfnamefont {Y.}~\bibnamefont {Silberberg}},\ }\bibfield  {title}
  {\bibinfo {title} {Entangled coherent states created by mixing squeezed
  vacuum and coherent light},\ }\href
  {https://doi.org/https://doi.org/10.1364/OPTICA.6.000753} {\bibfield
  {journal} {\bibinfo  {journal} {Optica}\ }\textbf {\bibinfo {volume} {6}},\
  \bibinfo {pages} {753} (\bibinfo {year} {2019})}\BibitemShut {NoStop}%
\bibitem [{\citenamefont {Wang}\ \emph {et~al.}(2016)\citenamefont {Wang},
  \citenamefont {Gao}, \citenamefont {Reinhold}, \citenamefont {Heeres},
  \citenamefont {Ofek}, \citenamefont {Chou}, \citenamefont {Axline},
  \citenamefont {Reagor}, \citenamefont {Blumoff}, \citenamefont {Sliwa},
  \citenamefont {Frunzio}, \citenamefont {Girvin}, \citenamefont {Jiang},
  \citenamefont {Mirrahimi}, \citenamefont {Devoret},\ and\ \citenamefont
  {Schoelkopf}}]{Wang2016}%
  \BibitemOpen
  \bibfield  {author} {\bibinfo {author} {\bibfnamefont {C.}~\bibnamefont
  {Wang}}, \bibinfo {author} {\bibfnamefont {Y.~Y.}\ \bibnamefont {Gao}},
  \bibinfo {author} {\bibfnamefont {P.}~\bibnamefont {Reinhold}}, \bibinfo
  {author} {\bibfnamefont {R.~W.}\ \bibnamefont {Heeres}}, \bibinfo {author}
  {\bibfnamefont {N.}~\bibnamefont {Ofek}}, \bibinfo {author} {\bibfnamefont
  {K.}~\bibnamefont {Chou}}, \bibinfo {author} {\bibfnamefont {C.}~\bibnamefont
  {Axline}}, \bibinfo {author} {\bibfnamefont {M.}~\bibnamefont {Reagor}},
  \bibinfo {author} {\bibfnamefont {J.}~\bibnamefont {Blumoff}}, \bibinfo
  {author} {\bibfnamefont {K.~M.}\ \bibnamefont {Sliwa}}, \bibinfo {author}
  {\bibfnamefont {L.}~\bibnamefont {Frunzio}}, \bibinfo {author} {\bibfnamefont
  {S.~M.}\ \bibnamefont {Girvin}}, \bibinfo {author} {\bibfnamefont
  {L.}~\bibnamefont {Jiang}}, \bibinfo {author} {\bibfnamefont
  {M.}~\bibnamefont {Mirrahimi}}, \bibinfo {author} {\bibfnamefont {M.~H.}\
  \bibnamefont {Devoret}},\ and\ \bibinfo {author} {\bibfnamefont {R.~J.}\
  \bibnamefont {Schoelkopf}},\ }\bibfield  {title} {\bibinfo {title} {A
  {S}chr{\"{o}}dinger cat living in two boxes},\ }\href
  {https://doi.org/https://doi.org/10.1126/science.aaf2941} {\bibfield
  {journal} {\bibinfo  {journal} {Science}\ }\textbf {\bibinfo {volume}
  {352}},\ \bibinfo {pages} {1087} (\bibinfo {year} {2016})}\BibitemShut
  {NoStop}%
\bibitem [{\citenamefont {M.~Kounalakis}\ and\ \citenamefont
  {Blanter}(2023)}]{Kounalakis2023}%
  \BibitemOpen
  \bibfield  {author} {\bibinfo {author} {\bibfnamefont {S.~V.}\ \bibnamefont
  {M.~Kounalakis}}\ and\ \bibinfo {author} {\bibfnamefont {Y.~M.}\ \bibnamefont
  {Blanter}},\ }\bibfield  {title} {\bibinfo {title} {Engineering entangled
  coherent states of magnons and phonons via a transmon qubit},\ }\href
  {https://doi.org/https://doi.org/10.1103/PhysRevB.108.224416} {\bibfield
  {journal} {\bibinfo  {journal} {Phys. Rev. B}\ }\textbf {\bibinfo {volume}
  {108}},\ \bibinfo {pages} {224416} (\bibinfo {year} {2023})}\BibitemShut
  {NoStop}%
\bibitem [{\citenamefont {H.~Jeon}\ and\ \citenamefont {Kim}(2024)}]{Jeon2024}%
  \BibitemOpen
  \bibfield  {author} {\bibinfo {author} {\bibfnamefont {J.~K. W. C. K.~K.}\
  \bibnamefont {H.~Jeon}, \bibfnamefont {J.~Kang}}\ and\ \bibinfo {author}
  {\bibfnamefont {T.}~\bibnamefont {Kim}},\ }\bibfield  {title} {\bibinfo
  {title} {Experimental realization of entangled coherent states in
  two-dimensional harmonic oscillators of a trapped ion},\ }\href
  {https://doi.org/https://doi.org/10.1103/PhysRevB.108.224416} {\bibfield
  {journal} {\bibinfo  {journal} {Sci. Rep.}\ }\textbf {\bibinfo {volume}
  {14}},\ \bibinfo {pages} {6847} (\bibinfo {year} {2024})}\BibitemShut
  {NoStop}%
\bibitem [{\citenamefont {Gerry}\ and\ \citenamefont
  {Campos}(2001)}]{Gerry2001}%
  \BibitemOpen
  \bibfield  {author} {\bibinfo {author} {\bibfnamefont {C.~C.}\ \bibnamefont
  {Gerry}}\ and\ \bibinfo {author} {\bibfnamefont {R.~A.}\ \bibnamefont
  {Campos}},\ }\bibfield  {title} {\bibinfo {title} {Generation of maximally
  entangled photonic states with a quantum-optical {Fredkin} gate},\ }\href
  {https://doi.org/https://doi.org/10.1103/PhysRevA.64.063814} {\bibfield
  {journal} {\bibinfo  {journal} {Phys. Rev. A}\ }\textbf {\bibinfo {volume}
  {64}},\ \bibinfo {pages} {063814} (\bibinfo {year} {2001})}\BibitemShut
  {NoStop}%
\bibitem [{\citenamefont {Kok}\ \emph {et~al.}(2002)\citenamefont {Kok},
  \citenamefont {Lee},\ and\ \citenamefont {Dowling}}]{Kok2002}%
  \BibitemOpen
  \bibfield  {author} {\bibinfo {author} {\bibfnamefont {P.}~\bibnamefont
  {Kok}}, \bibinfo {author} {\bibfnamefont {H.}~\bibnamefont {Lee}},\ and\
  \bibinfo {author} {\bibfnamefont {J.~P.}\ \bibnamefont {Dowling}},\
  }\bibfield  {title} {\bibinfo {title} {Creation of large photon-number path
  entanglement conditioned on photodetection},\ }\href
  {https://doi.org/https://doi.org/10.1103/PhysRevA.65.052104} {\bibfield
  {journal} {\bibinfo  {journal} {Phys. Rev. A}\ }\textbf {\bibinfo {volume}
  {65}},\ \bibinfo {pages} {052104} (\bibinfo {year} {2002})}\BibitemShut
  {NoStop}%
\bibitem [{\citenamefont {Pryde}\ and\ \citenamefont
  {White}(2003)}]{Pryde2003}%
  \BibitemOpen
  \bibfield  {author} {\bibinfo {author} {\bibfnamefont {G.~J.}\ \bibnamefont
  {Pryde}}\ and\ \bibinfo {author} {\bibfnamefont {A.~G.}\ \bibnamefont
  {White}},\ }\bibfield  {title} {\bibinfo {title} {Creation of maximally
  entangled photon-number states using optical fiber multiports},\ }\href
  {https://doi.org/https://doi.org/10.1103/PhysRevA.68.052315} {\bibfield
  {journal} {\bibinfo  {journal} {Phys. Rev. A}\ }\textbf {\bibinfo {volume}
  {68}},\ \bibinfo {pages} {052315} (\bibinfo {year} {2003})}\BibitemShut
  {NoStop}%
\bibitem [{\citenamefont {Walther}\ \emph {et~al.}(2004)\citenamefont
  {Walther}, \citenamefont {Pan}, \citenamefont {Aspelmeyer}, \citenamefont
  {Ursin}, \citenamefont {Gasparoni},\ and\ \citenamefont
  {Zeilinger}}]{Walther2004}%
  \BibitemOpen
  \bibfield  {author} {\bibinfo {author} {\bibfnamefont {P.}~\bibnamefont
  {Walther}}, \bibinfo {author} {\bibfnamefont {J.-W.}\ \bibnamefont {Pan}},
  \bibinfo {author} {\bibfnamefont {M.}~\bibnamefont {Aspelmeyer}}, \bibinfo
  {author} {\bibfnamefont {R.}~\bibnamefont {Ursin}}, \bibinfo {author}
  {\bibfnamefont {S.}~\bibnamefont {Gasparoni}},\ and\ \bibinfo {author}
  {\bibfnamefont {A.}~\bibnamefont {Zeilinger}},\ }\bibfield  {title} {\bibinfo
  {title} {De {Broglie} wavelength of a non-local four-photon state},\ }\href
  {https://doi.org/https://doi.org/10.1038/nature02552} {\bibfield  {journal}
  {\bibinfo  {journal} {Nature}\ }\textbf {\bibinfo {volume} {429}},\ \bibinfo
  {pages} {158} (\bibinfo {year} {2004})}\BibitemShut {NoStop}%
\bibitem [{\citenamefont {Mitchell}\ \emph {et~al.}(2004)\citenamefont
  {Mitchell}, \citenamefont {Lundeen},\ and\ \citenamefont
  {Steinberg}}]{Mitchell2004}%
  \BibitemOpen
  \bibfield  {author} {\bibinfo {author} {\bibfnamefont {M.~W.}\ \bibnamefont
  {Mitchell}}, \bibinfo {author} {\bibfnamefont {J.~S.}\ \bibnamefont
  {Lundeen}},\ and\ \bibinfo {author} {\bibfnamefont {A.~M.}\ \bibnamefont
  {Steinberg}},\ }\bibfield  {title} {\bibinfo {title} {Super-resolving phase
  measurements with a multiphoton entangled state},\ }\href
  {https://doi.org/https://doi.org/10.1038/nature02493} {\bibfield  {journal}
  {\bibinfo  {journal} {Nature}\ }\textbf {\bibinfo {volume} {429}},\ \bibinfo
  {pages} {161} (\bibinfo {year} {2004})}\BibitemShut {NoStop}%
\bibitem [{\citenamefont {Resch}\ \emph {et~al.}(2007)\citenamefont {Resch},
  \citenamefont {Pregnell}, \citenamefont {Prevedel}, \citenamefont
  {Gilchrist}, \citenamefont {Pryde}, \citenamefont {{O'Brien}},\ and\
  \citenamefont {White}}]{Resch2007}%
  \BibitemOpen
  \bibfield  {author} {\bibinfo {author} {\bibfnamefont {K.~J.}\ \bibnamefont
  {Resch}}, \bibinfo {author} {\bibfnamefont {K.~L.}\ \bibnamefont {Pregnell}},
  \bibinfo {author} {\bibfnamefont {R.}~\bibnamefont {Prevedel}}, \bibinfo
  {author} {\bibfnamefont {A.}~\bibnamefont {Gilchrist}}, \bibinfo {author}
  {\bibfnamefont {G.~J.}\ \bibnamefont {Pryde}}, \bibinfo {author}
  {\bibfnamefont {J.~L.}\ \bibnamefont {{O'Brien}}},\ and\ \bibinfo {author}
  {\bibfnamefont {A.~G.}\ \bibnamefont {White}},\ }\bibfield  {title} {\bibinfo
  {title} {Time-reversal and super-resolving phase measurements},\ }\href
  {https://doi.org/https://doi.org/10.1103/PhysRevLett.98.223601} {\bibfield
  {journal} {\bibinfo  {journal} {Phys. Rev. Lett.}\ }\textbf {\bibinfo
  {volume} {98}},\ \bibinfo {pages} {223601} (\bibinfo {year}
  {2007})}\BibitemShut {NoStop}%
\bibitem [{\citenamefont {Cable}\ and\ \citenamefont
  {Dowling}(2007)}]{Cable2007}%
  \BibitemOpen
  \bibfield  {author} {\bibinfo {author} {\bibfnamefont {H.}~\bibnamefont
  {Cable}}\ and\ \bibinfo {author} {\bibfnamefont {J.~P.}\ \bibnamefont
  {Dowling}},\ }\bibfield  {title} {\bibinfo {title} {Efficient generation of
  large number-path entanglement using only linear optics and feed-forward},\
  }\href {https://doi.org/https://doi.org/10.1103/PhysRevLett.99.163604}
  {\bibfield  {journal} {\bibinfo  {journal} {Phys. Rev. Lett.}\ }\textbf
  {\bibinfo {volume} {99}},\ \bibinfo {pages} {163604} (\bibinfo {year}
  {2007})}\BibitemShut {NoStop}%
\bibitem [{\citenamefont {Hofmann}\ and\ \citenamefont
  {Ono}(2007)}]{Hofmann2007}%
  \BibitemOpen
  \bibfield  {author} {\bibinfo {author} {\bibfnamefont {H.~F.}\ \bibnamefont
  {Hofmann}}\ and\ \bibinfo {author} {\bibfnamefont {T.}~\bibnamefont {Ono}},\
  }\bibfield  {title} {\bibinfo {title} {High-photon-number path entanglement
  in the interference of spontaneously down-converted photon pairs with
  coherent laser light},\ }\href
  {https://doi.org/https://doi.org/10.1103/PhysRevA.76.031806} {\bibfield
  {journal} {\bibinfo  {journal} {Phys. Rev. A}\ }\textbf {\bibinfo {volume}
  {76}},\ \bibinfo {pages} {031806(R)} (\bibinfo {year} {2007})}\BibitemShut
  {NoStop}%
\bibitem [{\citenamefont {Jones}\ \emph {et~al.}(2009)\citenamefont {Jones},
  \citenamefont {Karlen}, \citenamefont {Fitzsimons}, \citenamefont {Ardavan},
  \citenamefont {Benjamin}, \citenamefont {Briggs},\ and\ \citenamefont
  {Morton}}]{Jones2009}%
  \BibitemOpen
  \bibfield  {author} {\bibinfo {author} {\bibfnamefont {J.~A.}\ \bibnamefont
  {Jones}}, \bibinfo {author} {\bibfnamefont {S.~D.}\ \bibnamefont {Karlen}},
  \bibinfo {author} {\bibfnamefont {J.}~\bibnamefont {Fitzsimons}}, \bibinfo
  {author} {\bibfnamefont {A.}~\bibnamefont {Ardavan}}, \bibinfo {author}
  {\bibfnamefont {S.~C.}\ \bibnamefont {Benjamin}}, \bibinfo {author}
  {\bibfnamefont {G.~A.~D.}\ \bibnamefont {Briggs}},\ and\ \bibinfo {author}
  {\bibfnamefont {J.~J.~L.}\ \bibnamefont {Morton}},\ }\bibfield  {title}
  {\bibinfo {title} {Magnetic field sensing beyond the standard quantum limit
  using 10-spin {NOON} states},\ }\href
  {https://doi.org/https://doi.org/10.1126/science.1170730} {\bibfield
  {journal} {\bibinfo  {journal} {Science}\ }\textbf {\bibinfo {volume}
  {324}},\ \bibinfo {pages} {1166} (\bibinfo {year} {2009})}\BibitemShut
  {NoStop}%
\bibitem [{\citenamefont {Afek}\ \emph {et~al.}(2010)\citenamefont {Afek},
  \citenamefont {Ambar},\ and\ \citenamefont {Silberberg}}]{Afek2010}%
  \BibitemOpen
  \bibfield  {author} {\bibinfo {author} {\bibfnamefont {I.}~\bibnamefont
  {Afek}}, \bibinfo {author} {\bibfnamefont {O.}~\bibnamefont {Ambar}},\ and\
  \bibinfo {author} {\bibfnamefont {Y.}~\bibnamefont {Silberberg}},\ }\bibfield
   {title} {\bibinfo {title} {High-{NOON} states by mixing quantum and
  classical light},\ }\href
  {https://doi.org/https://doi.org/10.1126/science.1188172} {\bibfield
  {journal} {\bibinfo  {journal} {Science}\ }\textbf {\bibinfo {volume}
  {328}},\ \bibinfo {pages} {879} (\bibinfo {year} {2010})}\BibitemShut
  {NoStop}%
\bibitem [{\citenamefont {Merkel}\ and\ \citenamefont
  {Wilhelm}(2010)}]{Merkel2010}%
  \BibitemOpen
  \bibfield  {author} {\bibinfo {author} {\bibfnamefont {S.~T.}\ \bibnamefont
  {Merkel}}\ and\ \bibinfo {author} {\bibfnamefont {F.~K.}\ \bibnamefont
  {Wilhelm}},\ }\bibfield  {title} {\bibinfo {title} {Generation and detection
  of {NOON} states in superconducting circuits},\ }\href
  {https://doi.org/https://doi.org/10.1088/1367-2630/12/9/093036} {\bibfield
  {journal} {\bibinfo  {journal} {New J. Phys.}\ }\textbf {\bibinfo {volume}
  {12}},\ \bibinfo {pages} {093036} (\bibinfo {year} {2010})}\BibitemShut
  {NoStop}%
\bibitem [{\citenamefont {Wang}\ \emph {et~al.}(2011)\citenamefont {Wang},
  \citenamefont {Mariantoni}, \citenamefont {Bialczak}, \citenamefont
  {Lenander}, \citenamefont {Lucero}, \citenamefont {Neeley}, \citenamefont
  {{O'Connell}}, \citenamefont {Sank}, \citenamefont {Weides}, \citenamefont
  {Wenner}, \citenamefont {Yamamoto}, \citenamefont {Yin}, \citenamefont
  {Zhao}, \citenamefont {Martinis},\ and\ \citenamefont {Cleland}}]{Wang2011}%
  \BibitemOpen
  \bibfield  {author} {\bibinfo {author} {\bibfnamefont {H.}~\bibnamefont
  {Wang}}, \bibinfo {author} {\bibfnamefont {M.}~\bibnamefont {Mariantoni}},
  \bibinfo {author} {\bibfnamefont {R.~C.}\ \bibnamefont {Bialczak}}, \bibinfo
  {author} {\bibfnamefont {M.}~\bibnamefont {Lenander}}, \bibinfo {author}
  {\bibfnamefont {E.}~\bibnamefont {Lucero}}, \bibinfo {author} {\bibfnamefont
  {M.}~\bibnamefont {Neeley}}, \bibinfo {author} {\bibfnamefont {A.~D.}\
  \bibnamefont {{O'Connell}}}, \bibinfo {author} {\bibfnamefont
  {D.}~\bibnamefont {Sank}}, \bibinfo {author} {\bibfnamefont {M.}~\bibnamefont
  {Weides}}, \bibinfo {author} {\bibfnamefont {J.}~\bibnamefont {Wenner}},
  \bibinfo {author} {\bibfnamefont {T.}~\bibnamefont {Yamamoto}}, \bibinfo
  {author} {\bibfnamefont {Y.}~\bibnamefont {Yin}}, \bibinfo {author}
  {\bibfnamefont {J.}~\bibnamefont {Zhao}}, \bibinfo {author} {\bibfnamefont
  {J.~M.}\ \bibnamefont {Martinis}},\ and\ \bibinfo {author} {\bibfnamefont
  {A.~N.}\ \bibnamefont {Cleland}},\ }\bibfield  {title} {\bibinfo {title}
  {Deterministic entanglement of photons in two superconducting microwave
  resonators},\ }\href
  {https://doi.org/https://doi.org/10.1103/PhysRevLett.106.060401} {\bibfield
  {journal} {\bibinfo  {journal} {Phys. Rev. Lett.}\ }\textbf {\bibinfo
  {volume} {106}},\ \bibinfo {pages} {060401} (\bibinfo {year}
  {2011})}\BibitemShut {NoStop}%
\bibitem [{\citenamefont {Zhang}\ \emph {et~al.}(2018)\citenamefont {Zhang},
  \citenamefont {Um}, \citenamefont {Lv}, \citenamefont {Zhang}, \citenamefont
  {Duan},\ and\ \citenamefont {Kim}}]{Zhang2018}%
  \BibitemOpen
  \bibfield  {author} {\bibinfo {author} {\bibfnamefont {J.}~\bibnamefont
  {Zhang}}, \bibinfo {author} {\bibfnamefont {M.}~\bibnamefont {Um}}, \bibinfo
  {author} {\bibfnamefont {D.}~\bibnamefont {Lv}}, \bibinfo {author}
  {\bibfnamefont {J.-N.}\ \bibnamefont {Zhang}}, \bibinfo {author}
  {\bibfnamefont {L.-M.}\ \bibnamefont {Duan}},\ and\ \bibinfo {author}
  {\bibfnamefont {K.}~\bibnamefont {Kim}},\ }\bibfield  {title} {\bibinfo
  {title} {{NOON} states of nine quantized vibrations in two radial modes of a
  trapped ion},\ }\href
  {https://doi.org/https://doi.org/10.1103/PhysRevLett.121.160502} {\bibfield
  {journal} {\bibinfo  {journal} {Phys. Rev. Lett.}\ }\textbf {\bibinfo
  {volume} {121}},\ \bibinfo {pages} {160502} (\bibinfo {year}
  {2018})}\BibitemShut {NoStop}%
\bibitem [{\citenamefont {Z{\'u}{\~n}iga-Segundo}\ \emph
  {et~al.}(2013)\citenamefont {Z{\'u}{\~n}iga-Segundo}, \citenamefont
  {Ju{\'a}rez-Amaro}, \citenamefont {Soto-Eguibar},\ and\ \citenamefont
  {Moya-Cessa}}]{ZunigaSegundo2013}%
  \BibitemOpen
  \bibfield  {author} {\bibinfo {author} {\bibfnamefont {A.}~\bibnamefont
  {Z{\'u}{\~n}iga-Segundo}}, \bibinfo {author} {\bibfnamefont {R.}~\bibnamefont
  {Ju{\'a}rez-Amaro}}, \bibinfo {author} {\bibfnamefont {F.}~\bibnamefont
  {Soto-Eguibar}},\ and\ \bibinfo {author} {\bibfnamefont {H.~M.}\ \bibnamefont
  {Moya-Cessa}},\ }\bibfield  {title} {\bibinfo {title} {Generation of {MOON}
  states in ion-laser interactions},\ }\href
  {https://doi.org/http://dx.doi.org/10.12785/qir/010103} {\bibfield  {journal}
  {\bibinfo  {journal} {Quant. Inf. Rev.}\ }\textbf {\bibinfo {volume} {1}},\
  \bibinfo {pages} {19} (\bibinfo {year} {2013})}\BibitemShut {NoStop}%
\bibitem [{\citenamefont {Puebla}\ \emph {et~al.}(2019)\citenamefont {Puebla},
  \citenamefont {Casanova}, \citenamefont {Houhou}, \citenamefont {Solano},\
  and\ \citenamefont {Paternostro}}]{Puebla2019}%
  \BibitemOpen
  \bibfield  {author} {\bibinfo {author} {\bibfnamefont {R.}~\bibnamefont
  {Puebla}}, \bibinfo {author} {\bibfnamefont {J.}~\bibnamefont {Casanova}},
  \bibinfo {author} {\bibfnamefont {O.}~\bibnamefont {Houhou}}, \bibinfo
  {author} {\bibfnamefont {E.}~\bibnamefont {Solano}},\ and\ \bibinfo {author}
  {\bibfnamefont {M.}~\bibnamefont {Paternostro}},\ }\bibfield  {title}
  {\bibinfo {title} {Quantum simulation of multiphoton and nonlinear
  dissipative spin-boson models},\ }\href
  {https://doi.org/https://doi.org/10.1103/PhysRevA.99.032303} {\bibfield
  {journal} {\bibinfo  {journal} {Phys. Rev. A}\ }\textbf {\bibinfo {volume}
  {99}},\ \bibinfo {pages} {032303} (\bibinfo {year} {2019})}\BibitemShut
  {NoStop}%
\bibitem [{\citenamefont {Magnus}(1954)}]{Magnus1954}%
  \BibitemOpen
  \bibfield  {author} {\bibinfo {author} {\bibfnamefont {W.}~\bibnamefont
  {Magnus}},\ }\bibfield  {title} {\bibinfo {title} {On the exponential
  solution of differential equations for a linear operator},\ }\href
  {https://doi.org/https://doi.org/10.1002/cpa.3160070404} {\bibfield
  {journal} {\bibinfo  {journal} {Comm. Pure App. Math.}\ }\textbf {\bibinfo
  {volume} {7}},\ \bibinfo {pages} {649} (\bibinfo {year} {1954})}\BibitemShut
  {NoStop}%
\bibitem [{\citenamefont {Plenio}(2005)}]{Plenio2005}%
  \BibitemOpen
  \bibfield  {author} {\bibinfo {author} {\bibfnamefont {M.~B.}\ \bibnamefont
  {Plenio}},\ }\bibfield  {title} {\bibinfo {title} {Logarithmic negativity: A
  full entanglement monotone that is not convex},\ }\href
  {https://doi.org/10.1103/PhysRevLett.95.090503} {\bibfield  {journal}
  {\bibinfo  {journal} {Phys. Rev. Lett.}\ }\textbf {\bibinfo {volume} {95}},\
  \bibinfo {pages} {090503} (\bibinfo {year} {2005})}\BibitemShut {NoStop}%
\bibitem [{\citenamefont {Q.-P.~Su}\ and\ \citenamefont
  {Zheng}(2014)}]{Su2014}%
  \BibitemOpen
  \bibfield  {author} {\bibinfo {author} {\bibfnamefont {C.-P.~Y.}\
  \bibnamefont {Q.-P.~Su}}\ and\ \bibinfo {author} {\bibfnamefont {S.-B.}\
  \bibnamefont {Zheng}},\ }\bibfield  {title} {\bibinfo {title} {Fast and
  simple scheme for generating {NOON} states of photons in circuit {QED}},\
  }\href {https://doi.org/https://doi.org/10.1038/srep03898} {\bibfield
  {journal} {\bibinfo  {journal} {Sci. Rep.}\ }\textbf {\bibinfo {volume}
  {4}},\ \bibinfo {pages} {3898} (\bibinfo {year} {2014})}\BibitemShut
  {NoStop}%
\bibitem [{\citenamefont {J.~Grebel}\ and\ \citenamefont
  {Cleland}(2024)}]{Grebel2024}%
  \BibitemOpen
  \bibfield  {author} {\bibinfo {author} {\bibfnamefont {M.-H. C. G. A. C. R.
  C. Y. J. J. J. M. M. R. G. P. H. Q. X.~W.}\ \bibnamefont {J.~Grebel},
  \bibfnamefont {H.~Yan}}\ and\ \bibinfo {author} {\bibfnamefont {A.~N.}\
  \bibnamefont {Cleland}},\ }\bibfield  {title} {\bibinfo {title}
  {Bidirectional multiphoton communication between remote superconducting
  nodes},\ }\href
  {https://doi.org/https://doi.org/10.1103/PhysRevLett.132.047001} {\bibfield
  {journal} {\bibinfo  {journal} {Phys. Rev. Lett.}\ }\textbf {\bibinfo
  {volume} {132}},\ \bibinfo {pages} {047001} (\bibinfo {year}
  {2024})}\BibitemShut {NoStop}%
\end{thebibliography}%

\end{document}